\documentclass[12pt]{article}
\usepackage{amssymb,latexsym}
\usepackage{amsmath}

\columnwidth\textwidth

\begin{document}

\newcommand{\be}{\begin{equation}}
\newcommand{\ee}{\end{equation}}
\newcommand{\beann}{\begin{eqnarray*}}
\newcommand{\eeann}{\end{eqnarray*}}
\newcommand{\bea}{\begin{eqnarray}}
\newcommand{\eea}{\end{eqnarray}}
\newcommand{\nn}{\nonumber}
\newtheorem{df}{Definition}
\newtheorem{thm}{Theorem}
\newtheorem{lem}{Lemma}

\begin{titlepage}

\noindent
\hspace*{11cm} BUTP-02/4\\
\vspace*{1cm}
\begin{center}
{\LARGE Quantum theory of gravitational collapse \\
 (lecture notes on quantum conchology)}

\vspace{2cm}

P. H\'{a}j\'{\i}\v{c}ek \\
Institute for Theoretical Physics \\
University of Bern \\
Sidlerstrasse 5, CH-3012 Bern, Switzerland

\vspace{1.5cm}

March 2002 \\ \vspace*{1cm}

\nopagebreak[4]

\begin{abstract}
  Preliminary version No.~2 of the lecture notes for the talk ``Quantum theory
  of gravitational collapse'' given at the 271. WE-Heraeus-Seminar ``Aspects of
  Quantum Gravity'' at Bad Honnef, 25 February---1 March 2002.
\end{abstract}

\end{center}

\end{titlepage}

\section{Introduction}
The issue of gravitational collapse includes not only the theory of black
holes and of their classical and quantum properties, but also the serious
problem of singularities. We believe that the singularities can be cured by
quantum theory, but not by any semiclassical approximation thereof. The
classical solutions near the singularity has to be changed strongly. New ideas
concerning singularities are likely to influence the theory of black holes,
too.

The rejection of WKB approximations forces us to another kind of approximative
schema. Simplified models constitute a promissing alternative.  Our method of
dealing with the problem will, therefore, employ simplified models and a kind
of effective theory of gravity. It does not worry about the final form of a
full-fledged theory of quantum gravity. This need not be completely
unreasonable approach. Even if the ultimate quantum gravity theory were known,
most calculations would still be performed by an approximation schema within
some effective theory and for simplified models, such as the method of
effective quantum theory in QCD. Another example of the method of simplified
models used successfully today is the theory of atomic spectra, which belongs,
strictly speaking, to the field of quantum electrodynamics.  Still, infinite
number of degrees of freedom can be safely frozen and Schroedinger equation
applied.  Such methods can give useful hints for the gravitational collapse
also because of the fact that the mainstay of the black hole geometry is
formed by the purely dependent degrees of freedom of the gravitational field
(uniqueness theorems!), and these degrees of freedom have no proper quantum
character of their own.

The model we work with is a spherically symmetric thin shell. Historically,
the first physicist to use shells in quantum theory seems to be Dirac
\cite{Dconch}.  We shall try to do justice to the model and to take it
absolutely seriously.  There is no point in rushing towards prejudiced
conceptions disregarding all kinds of problems that may emerge on the way.

These lecture notes are to explain in a coherent way a number of ideas
scattered in various papers but they also contain some new results.
Sec.~\ref{sec:gauge} gives a general account of some gauge-invariant methods
in the canonical theory of generally covariant systems, whose spiritual
fathers are Dirac, Bergmann, Kucha\v{r} and Rovelli.  New seems to be the
observation that many conceptual and practical problems admit a solution if
one restrict oneself to asymptotically flat models. This is a natural
framework for the gravitational collapse. The asymptotic structure defines a
decomposition of the full diffeomorphism group into the gauge group, the group
of asymptotic symmetries and the rest that can be ignored.  A physically
meaningful choice of a complete set of gauge-invariant quantities---the
so-called Dirac observables---is also provided by the structure. The
asymptotic symmetries determine a gauge-invariant dynamics. In this way, some
aspects of the notorious problem of time are dealt with.

Sec.~\ref{sec:model} starts with an application of the methods to the model of
a single, spherically symmetric, thin shell of light-like matter surrounded by
its own gravitational field. The study of the space of the corresponding
solutions to Einstein equations leads to the identification of a complete set
of Dirac observables. Thus, the basic observables are found independently of
the asymptotic structure in this (lucky) case, but they can still be
considered as asymptotic properties of the system. The canonical theory of the
shell due to Louko, Whiting and Friedman is utilized to calculate the Poisson
brackets of the observables and to find the generators of the asymptotic
symmetries. The main new point is a careful description of asymptotic frames.

Finally, a quantum theory of the model is constructed in
Sec.~\ref{sec:quantum}. The method of choice is the group-theoretical
quantization because the phase space has a complicated boundary and the group
quantization has been invented to deal with such a situation. The results are
rather surprising: Some quantum shells contract, cross their Schwarzschild
radius inwards, bounce at the centre without creating a singularity, cross
their Schwarzschild radius outwards, and re-expands to the infinity from which
they originally came. This is no joke! It is possible because the quantum
Schwarzschild radius is a mixture of black and white hole states. It is
amusing that an old calculation \cite{F-V} based on a very different technique
has come to somewhat similar results. A couple of ideas are listed of how
these results could be reconciled with the ``observational evidence for black
holes'' in astrophysics. An interesting problem for future research arises:
what is the nature of quantum geometry?

\section{Gauge-invariant method in canonical theory \\
of generally covariant systems}
\label{sec:gauge}
The present section attempts to give a general account of some new methods
that have been developed in a couple of recent years. It is also a description
of a project. It contains many claims without proofs. Those of these claims
that make sense for finite dimensional systems have been proved and discussed
extensively in \cite{H1} and \cite{H2} for general finite dimensional systems.
The rest of the claims without proofs have hitherto been proved only for a few
special cases, one of which will be described in the next section.

\subsection{Space of solutions, gauge group \\ and asymptotic symmetries}
\label{sec:sol}
Mathematically, any {\em generally covariant system} is constituted by a set
$\Psi$ of geometrical objects (fields, and submanifolds such as particles or
shells) on a background manifold $\mathcal M$. The set $\Psi$ always includes
a Lorentzian metric $g$ on $\mathcal M$; it is sufficient to consider $\Psi$
as consisting just of $g$ in order to understand all manipulations with $\Psi$
in this lecture. Details about complicated systems are explained in
\cite{H-Kij} and \cite{H-Kie}. The background $\mathcal M$ has just a
topological and differential structure (bare manifold without boundary) and
has the meaning of a {\em spacetime} manifold. It can be of any dimension, but
we assume here that it is four-dimensional. Several different backgrounds can
be needed for one generally covariant system. For example, all
three-dimensional manifolds that admit a Riemannian metric can play a role of
Cauchy surfaces in general relativity; different topologies of Cauchy surface
define different background manifolds.

The object $\Psi$ has to satisfy a dynamical equation on $\mathcal M$ (such as
Einstein equations for $g$). We allow for a domain ${\mathcal D}_\Psi$ of
$\Psi$ in ${\mathcal M}$ where the solution is well defined to be a proper
subset of ${\mathcal M}$. General covariance is the following property. Let
Diff$\mathcal M$ be the diffeomorphism group of $\mathcal M$ and let
$({\mathcal M},\Psi)$ be a spacetime that solves the dynamical equation. Then
$({\mathcal M},\varphi_*\Psi)$ is another solution for any $\varphi \in
\text{Diff}{\mathcal M}$. The symbol $\varphi_*$ denotes the push forward of
all geometrical objects collected within $\Psi$ by $\varphi$. For example, let
$\mathcal M$ admits a global coordinate chart and let $\{X^\mu\}$ denote some
global coordinates on $\mathcal M$; the action of a diffeomorphism $\varphi$
on a metric field $g_{\mu\nu}(X)$ has the following result:
\[
  (\varphi_*g)_{\mu\nu}(X) = g_{\rho\sigma}\bigl(Y(X)\bigr) \frac{\partial
  Y^\rho}{\partial X^\mu}\frac{\partial Y^\sigma}{\partial X^\nu}\ ,
\]
where the functions $Y^\rho(X)$ are defined by $\varphi$ and define the action
of $\varphi$ in terms of the coordinates $\{X^\mu\}$:
\[
  Y^\rho(X) = X^\rho\bigl(\varphi^{-1}(X)\bigr)\ .
\]

The map $\varphi_*$ is an ``active'' transformation so that e.g.\ $\varphi_*g$
is a different field on $\mathcal M$ than $g$ is, in general. If $\Psi$ and
$\varphi$ are such that $\varphi_*\Psi = \Psi$, then we say that $\Psi$ admits
a {\em spacetime symmetry} $\varphi$. The reason why we prefer to work with
active transformations rather than coordinate transformations is a technical
one: active transformtions can be globally defined even if there are no global
coordinate charts. In consequence of that, they form groups so that the whole
powerful apparatus of group theory can be applied.

We shall restrict ourselves to the special case of systems with {\em
  asymptotically flat} solutions. There are coordinate free methods (due to
Penrose) to decide whether or not a spacetime is asymptotically flat (see,
e.g., \cite{H-E} or \cite{C-J-K}). Then there is a universal {\em asymptotic
  boundary} $\partial_{\text{as}}{\mathcal M}$ carrying an {\em asymptotic
  structure} that can be considered as common to all such solutions. For
example, asymptotically flat solutions to Einstein equations all have the same
scri, which has a geometrical structure that is richer than just the
topological and differential one. This structure posses a symmetry group that
we call $G_{\text{as}}$\footnote{For example, a symmetry of the geometrical
  structure of flat spacetime is an isometry of the spacetime; the symmetry
  group of it is the Poincar\'{e} group.}.  We shall not go into detail here,
but we shall give a more complete account for an example later. The asymptotic
structure enables to select {\em standard asymptotical frames of reference} at
$\partial_{\text{as}}{\mathcal M}$ that are also common to all solutions and
are inhabitated by {\em asymptotic observers}.

One can try to give a more precise account of these ideas as follows. Let
$\bar{\Gamma}$ be the space of all asymptotically flat solutions on $\mathcal
M$. We assume that there is a subspace $\tilde{\Gamma} \subset \bar{\Gamma}$
and a subgroup $G$ of Diff$\mathcal M$ with the following properties.
\begin{enumerate}
\item For each $\Psi \in \tilde{\Gamma}$, the spacetime $({\mathcal M},\Psi)$
  can be conformally extended to $(\bar{\mathcal M}_\Psi,\Psi)$ by attaching a
  scri ${\mathcal I}_\Psi$ to $\mathcal M$ and a neighbourhood
  $\tilde{\mathcal I}_\Psi$ of ${\mathcal I}_\Psi$ in $\bar{\mathcal M}_\Psi$.
  Then there is a diffeomorphism $\Phi_\Psi : \tilde{\mathcal I}_\Psi \mapsto
  \bar{\mathcal M}_0$, such that $\Phi_\Psi$ maps ${\mathcal I}_\Psi$ onto
  $\mathcal I$, where $\bar{\mathcal M}_0$ is the conformal compactification
  of Minkowski spacetime and $\mathcal I$ is the scri of $\bar{\mathcal M}_0$.
\item Let $\varphi \in G$, $\Psi \in \tilde{\Gamma}$, $\Psi' = \varphi_*\Psi$
  and $U = \Phi_\Psi\tilde{\mathcal I}_\Psi \cap
  \Phi_{\Psi'}\tilde{\mathcal I}_{\Psi'}$. Then $\varphi$ can be
  differentiably extended to ${\mathcal I}_\Psi$ and $\varphi_\Psi : U \mapsto
  U$ defined by $\varphi_\Psi = \Phi_{\Psi'} \circ \varphi \circ
  \Phi^{-1}_\Psi$ induces a map $\varphi_\Psi\vert_{\mathcal I}$ on $\mathcal
  I$ of Minkowski spacetime that preserves the asymptotic structure and is
  an element of the asymptotic symmetry group $G_{\text{as}}$,
  $\varphi_\Psi\vert_{\mathcal I} \in G_{\text{as}}$.
\item For given $\varphi \in G$, $\varphi_\Psi\vert_{\mathcal I}$ is
  independent of $\Psi$. In this way, there is a well-defined map
  $\pi_{\text{as}} : G \mapsto G_{\text{as}}$, and we assume that
  $\pi_{\text{as}}$ is onto. It follows that this map is a homeomorphism if it
  exists. Let us define the {\em gauge group} $G_0$ of the system to be the
  kernel of $\pi_{\text{as}}$. It follows that $G_0$ is a normal subgroup of
  $G$.
\item For each $\varphi \in G_0$ and $\Psi \in \tilde{\Gamma}$, the extension
  of $\varphi$ to ${\mathcal I}_\Psi$ is an odd-parity
  supertranslation\footnote{Justification for this assumption must be taken
    from the canonical theory of the next subsection; this shows how tentative
    and preliminary status the present subsection has.} on ${\mathcal
    I}_\Psi$.
\end{enumerate}
If this procedure works, then it is surely non unique. For example, $\mathcal
I$ can be chosen to lie in a ``different corner'' of $\mathcal M$. We suppose
that the non-uniqueness can either be limited by some suitable additional
requirements or that it will not manifest itself in physical results.

The quotient $\Gamma := \tilde{\Gamma}/G_0$ is the {\em physical phase space}.
(To be a full-fledged phase space, it has to be equipped with a symplectic
structure.)  Let us denote the projection of the quotient by $\pi_{\text{sol}}
: \tilde{\Gamma} \mapsto \tilde{\Gamma}/G_0$.

Let $\varphi_1 \in G \setminus G_0$. Consider the action of $\varphi_1$ on a
point $\Psi_1$ of an orbit $\gamma$ of $G_0$ in $\tilde{\Gamma}$. Clearly,
$\varphi_{1*}\Psi$ lies in another orbit $\gamma'$ and $\gamma' \neq \gamma$.
Let $\Psi_2 \in \gamma$ and $\Psi_2 \neq \Psi_1$. Does $\varphi_{1*}\Psi_2$
lie in the same orbit as $\varphi_{1*}\Psi_1$? It does because $G_0$ is a
normal subgroup of $G$. Hence $\varphi_1$ sends orbits into orbits. Further,
let $\varphi_2$ be from the same class of $G/G_0$ as $\varphi_1$ is. Then
$\varphi_2$ also sends $\gamma$ into $\gamma'$. Since $G_{\text{as}}$ can be
identified with the factor group $G/G_0$, this shows that the asymptotic
symmetry group acts on the physical phase space.

Consider a surface $\sigma$ in $\tilde{\Gamma}$ that cuts each orbit exactly
once. Such a surface is called a {\em section} of the quotient
$\tilde{\Gamma}/G_0$. Each section of $\tilde{\Gamma}/G_0$ breaks the gauge
group and can, therefore, be called {\em covariant gauge fixing}. A section
$\sigma$ can also be described as follows. Let $X^\mu$ be coordinates on
$\mathcal M$ and $o^A$ coordinates on $\Gamma$. Then, for the special case
that $\Psi = g$, $\sigma$ determines the set of functions
\begin{equation}
  g_{\mu\nu}(o,X),
\label{sigma}
\end{equation}
that is, metric components with respect to the coordinates $X^\mu$ at the point
$g$ of intersection between $\sigma$ and the orbit given by the coordinates
$o^i$. The coordinates $X^\mu$ can be arbitrary and the same section can so be
represented by different sets of functions (\ref{sigma}). That is why the
gauge fixing is called ``covariant.'' Observe also that no global coordinate
chart is necessary to describe $\sigma$.

Transformations between two different gauge fixings is constituted by a set of
diffeomorphisms $\varphi(o)$, one for each orbit $\gamma$ determined by
coordinates $o^i$. Such gauge transformations are common in every gauge
theory. For example, in electrodynamics, one often uses the so-called Coulomb
gauge. This is defined by a differential equation for the components of the
potential. One also uses the so-called axial gauge that is defined by an
algebraic equation for the potential. The transformation between these two
gauges must, therefore, depend on the potential in a rather complicated and
non-local way. Quantum field theory constructed in a given gauge {\em cannot}
be made invariant with respect to all field dependent gauge transformations.
One hopes to overcome this difficulty by working with gauge-invariant
quantities.

Similarly, for generally covariant systems, the two quantum theories
constructed in two different gauges that are related by a field dependent
transformation are not unitarily equivalent, unless they are limited to
relations between Dirac observables \cite{paris}.

Let $o \in \Gamma$ and $\Psi(o) = \sigma \cap \pi_{\text{sol}}^{-1} o$.  It is
a simple exercise to show that there is an element $\varphi_o$ in each class
of $G/G_0$ such that $\varphi_{o*}\Psi(o) \in \sigma$; if $\Psi(o)$ does not
admit any spacetime symmetry, then $\varphi_o$ is even unique! Thus, given a
gauge fixing, we find a ($o$-dependent) extension of each asymptotic symmetry
to the whole of $\mathcal M$. Even if gauge (and, in general, $o$-) dependent,
this extension is a practical tool that will be used later.

\subsection{Phase space}
\label{sec:phs}
We assume now that $\mathcal M$ has the structure $\Sigma \times {\mathbb R}$
and that the dynamical equation admits a well-posed Cauchy problem on the
manifold $\Sigma$. Let us denote a Cauchy datum for the solution $\Psi$ by
$(\Psi_\Sigma,\Pi_\Sigma)$, where $\Psi_\Sigma$ and $\Pi_\Sigma$ are some
geometric objects on $\Sigma$. We assume that a Riemannian metric $^3g_{kl}$
with the meaning of the first fundamental form of $\Sigma$ belongs to
$\Psi_\Sigma$ and a symmetric tensor field $K_{kl}$ with the meaning of the
second fundamental form of $\Sigma$ belongs to $\Pi_\Sigma$. We assume further
that $(\Sigma,\ ^3g)$ is asymptotically Euclidean. This involves the existence
of a special coordinate patch in $\Sigma$ such that the corresponding
components of all objects within $\Psi_\Sigma$ and $\Pi_\Sigma$ satisfy
suitable fall-off asymptotic conditions, so-called {\em Cauchy data asymptotic
  conditions} (CDAC). Different examples of CDAC are given in \cite{B-O} and
in \cite{C-J-K}, Sec.~5.4.

For generally covariant systems, the objects $\Psi_\Sigma$ and $\Pi_\Sigma$
must satisfy some particular conditions (mostly differential equations) in
order to constitute an initial datum for a solutions. These equations are
called {\em constrains}.

The {\em phase space} $\mathcal P$ of the system is a manifold the points of
which are all unconstrained Cauchy data. It carries a symplectic structure
$\bar{\Omega}$, which defines Poisson brackets. We assume that such a space
can be constructed. For example, $^3g_{kl}$ and
\[
  \pi^{kl} := \sqrt{\text{Det}(^3g)}(^3g^{kl}\,^3g^{rs} -\ 
  ^3g^{kr}\,^3g^{ls})K_{rs} 
\]
are canonically conjugated variables for the case that $\Psi = g$ \cite{ADM}.

We assume further that all constrained data form a submanifold $\mathcal C$ of
$\mathcal P$, called {\em constraint surface}: each point of $\mathcal C$ is a
Cauchy datum for a solution and each Cauchy datum for a solution lies in
$\mathcal C$.

There is an important relation ``$\sim$'' between points of $\mathcal C$ that
can be defined as follows. The points $p_1$ and $p_2$ are said to satisfy $p_1
\sim p_2$ if the solutions $\Psi_1$ and $\Psi_2$ determined by them satisfy
\[
  \Psi_2 = \varphi_*\Psi_1
\]
for some $\varphi \in G_0$. We assume that ``$\sim$'' is an equivalence
relation.  (For general relativity, this has been shown in \cite{F-M}). The
equivalence class of points at $\mathcal C$ are called {\em c-orbits} and
denoted by $\lambda$. Each c-orbit determines a class of $G_0$-equivalent
solutions and vice versa. Hence, the quotient ${\mathcal C}/\lambda$ is the
physical phase space defined in Sec.~\ref{sec:sol}:
\[
  \Gamma = {\mathcal C}/\lambda.
\]
The quotient projection of the constraint surface $\mathcal C$ to the physical
phase space $\Gamma$ will be denoted by $\pi_{\text{phs}}$.

Let the constraint functionals
\[
  {\mathcal H}[\Psi_\Sigma,\Pi_\Sigma;x),\ {\mathcal
  H}_k[\Psi_\Sigma,\Pi_\Sigma;x), 
\]
where $k = 1,2,3$ and $x \in \Sigma$, define the constraint surface $\mathcal
C$ by ${\mathcal H} = 0$ and ${\mathcal H}_k = 0$. It seems that they can be
chosen for all generally covariant systems so that they obey the so-called
Dirac algebra \cite{teit}, see also \cite{koule}: Let $x^k$ be coordinates,
${\mathcal N}(x)$ a scalar and ${\mathcal N}^k(x)$ a vector fields on
$\Sigma$; the fields ${\mathcal N}(x)$ and ${\mathcal N}^k(x)$ are called {\em
  lapse} and {\em shift}, respectively. Let $\vec{\mathcal N} := ({\mathcal
  N},{\mathcal N}^1,{\mathcal N}^2,{\mathcal N}^3)$ and
\[
  {\mathcal H}[\vec{\mathcal N}] := \int_\Sigma d^3x\bigl({\mathcal
  N}(x){\mathcal H}(x) + {\mathcal N}^k(x){\mathcal H}_k(x)\bigr)\ .
\]
Then
\begin{equation}
  \left\{{\mathcal H}[\vec{\mathcal N}_1], {\mathcal H}[\vec{\mathcal
  N}_2]\right\} = {\mathcal H}[\vec{\mathcal N}]\ ,
\label{dirac1}
\end{equation}
where
\begin{eqnarray*}
  {\mathcal N} & = & {\mathcal N}_1^k \partial_k {\mathcal N}_2 - {\mathcal
  N}_2^k \partial_k {\mathcal N}_1\ , \\
  {\mathcal N}^k & = & {\mathcal N}_1^l \partial_l {\mathcal N}_2^k - {\mathcal
  N}_2^l \partial_l {\mathcal N}_1^k +\ ^3g^{kl}({\mathcal N}_1 \partial_l
  {\mathcal N}_2 - {\mathcal N}_2 \partial_l {\mathcal N}_1)\ .
\end{eqnarray*}
This implies that the Poisson brackets of the constraints vanish at the
constraint surface; such a constraint surface is called {\em first class}
\cite{D4}, \cite{H-T}. However, Eq.~(\ref{dirac1}) holds only if the lapse and
shift fields satisfy some fall-off conditions (see \cite{B-O}), called {\em
  gauge group asymptotic conditions} (GGAC). An unexpected result of
\cite{B-O} is that $\vec{\mathcal N}$ need not approach zero asymptotically in
order to satisfy (GGAC) but can contain arbitrary odd-parity
supertranslations.

The lapse and shift fields can even approach linear functions of coordinates
asymptotically. This corresponds to infinitesimal Poincar\'{e} transformations
at infinity (cf.~\cite{R-T} and \cite{B-O}). The condition that $\vec{N}
\rightarrow \vec{N}_\infty$, where $\vec{N}_\infty$ represents the asymptotic
behaviour of $\vec{N}$ corresponding to an element of the Lie algebra of
Poincar\'{e} group is called {\em Poincar\'{e} group asymptotic condition}
(PGAC); the form of $\vec{N}_\infty$ as a function on Poincar\'{e} algebra is
given in \cite{R-T} and \cite{B-O}. The corresponding functionals ${\mathcal
  H}[\vec{N}]$ are not differentiable (their variations lead to surface
integrals at infinity) or are not even convergent. This can be improved by
adding {\em surface terms} at infinity to them that we denote by ${\mathcal
  H}_\infty[\vec{N}_\infty]$: the functional form of ${\mathcal
  H}_\infty[\vec{N}_\infty]$ is given in \cite{R-T} and \cite{B-O}. An
analysis of similar surface terms for the null-infinity is given in
\cite{C-J-K}. The expressions ${\mathcal H}[\vec{N}] + {\mathcal
  H}_\infty[\vec{N}_\infty]$ have finite values on constraint surface and
generate asymptotic Poincar\'{e} transformation\footnote{The surface terms may
  be uniquely determined (except for additive constants) by boundary
  conditions imposed on the dynamics of the fields at the surface. For a
  finite surface, this has been shown by \cite{G-K}.}.

Observe that the CDAC form a part of the definition of the phase space, while
GGAC and PGAC help define some transformations on this phase space. In
particular, the CDAC must be preserved in all transformations satisfying the
GGAC or PGAC, see \cite{B-O}.  In this way, the set of all lapse and shift
fields is divided into those that are associated with infinitesimal gauge
transformations, infinitesimal dynamical symmetries and the rest that is not
interesting. Again, the gauge $\vec{N}$'s form a large subset of the symmetry
$\vec{N}$'s. What is the relation between this Cauchy surface picture and the
spacetime picture of subsection \ref{sec:sol}? We believe that such a relation
can be found using the technique invented by Kucha\v{r}, the so-called
Kucha\v{r} variables (embeddings, see subsection \ref{sec:trans}). In
particular, the CDAC must be related to asymptotic conditions imposed on the
embeddings. An example of how this may work will be given in the next section.

If $\vec{\mathcal N}$ satisfies the GGAC, the transformation generated by
${\mathcal H}[\vec{\mathcal N}]$ in the phase space shifts the points of
$\mathcal C$ along c-orbits, and any point of a given c-orbit can be reached
in this way from any other point of it. This is the reason to call $\lambda$ a
c-orbit (constraints orbit).

For any first class constraint surface, there is a unique way of how the
symplectic structure $\bar{\Omega}$ of the phase space $\mathcal P$ defines a
symplectic structure $\Omega$ on the physical phase space $\Gamma$: First step
is to pull back the two-form $\bar{\Omega}$ to $\mathcal C$. The result is a
two-form $\tilde{\Omega}$. The form $\tilde{\Omega}$ is degenerated along
c-orbits; it is not a symplectic form, but it is closed ($d\tilde{\Omega} =
0$, where $d$ denotes the external differentiation). The second step is a
proof that there is a unique symplectic form $\Omega$ on $\Gamma$ such that
the pull back of it by the projection $\pi_{\text{phs}}$ to $\mathcal C$ is
$\tilde{\Omega}$. In this way, we finally obtain the full-fledged physical
phase space $(\Gamma,\Omega)$.

\subsection{Observables and dynamical symmetries}
The action of the gauge group has been quotiented away from the physical phase
space, so it is a gauge-invariant structure. Functions on it are called {\em
  observables}:
\[
  o : \Gamma \mapsto {\mathbb R}\ .
\]
The central idea of the gauge-invariant method described by these lectures is
to answer all physically interesting questions by manipulating quantities
defined on the physical phase space, such as observables. To interpret or
justify such manipulations, we need the connection of $\Gamma$ to the larger
spaces $\mathcal C$, $\mathcal P$ and $\tilde{\Gamma}$. 

Each observable $o$ determines a function $\tilde{o}$ on $\mathcal C$ by
\begin{equation}
  \tilde{o} := o \circ \pi_{\text{phs}}\ .  
\label{defdir}
\end{equation}
Clearly, all such functions must be constant along c-orbits:
\[
  \left\{\tilde{o},{\mathcal H}[\vec{\mathcal N}]\right\} = 0
\]
for all $\vec{\mathcal N}$ that satisfy GGAC; they are called {\em Dirac
  observables} \cite{D4}. Examples of Dirac observable are the quantities
${\mathcal H}[\vec{N}] + {\mathcal H}_\infty[\vec{N}_\infty]$ for all fields
$\vec{N}$ that satisfy PGAC, see \cite{B-O}.

Let $\{o^i\}$ be some coordinates on $\Gamma$; they form a {\em complete set
  of observables}. Similarly, the corresponding functions $\tilde{o}^i$ form
{\em complete set of Dirac observables}. One needs complete sets with some
additional properties to quantize by a gauge-invariant way, and one needs
still more observables to describe all interesting properties of the quantum
system. This will be shown later by an example. The idea of basing
quantization of gravity on complete sets of Dirac observables encounters a
difficulty: there is no single quantity of this kind known in general
relativity, for example in the case when the Cauchy surface is compact
\cite{berg}, \cite{kuch}. Even if it were known, it would be non-local
\cite{torre}. In the asymptotically flat case, however, two complete sets are
known: they are associated with the asymptotic in- and out-fields. They have
been described, within a perturbation theory, by DeWitt (``asymptotic
invariants'') \cite{DW}, and for the exact theory by Ashtekar (``radiative
modes'') \cite{A}. We propose to work out a theory for this favorable case
first and then to see if anything can be done for the other cases.

Let $\bar{h} : {\mathcal P} \mapsto {\mathcal P}$ be a map that preserves the
symplectic form $\bar{\Omega}$ (symplectomorphism) and the constraint surface
$\mathcal C$ in ${\mathcal P}$. Such a map is called {\em extended dynamical
  symmetry} (extended to the extended phase space ${\mathcal P}$). We use the
word ``dynamical'' to distinguish it from spacetime symmetries.

Clearly, such a map $\bar{h}$ induces a map of the constraint surface onto
itself: $\tilde{h} := \bar{h}\vert_{\mathcal C}$ and $\tilde{h} : {\mathcal C}
\mapsto {\mathcal C}$. The most important properties of $\tilde{h}$ are: it
preserves the form $\tilde{\Omega}$ and maps the c-orbits onto c-orbits. Such
a map $\tilde{h}$ is called {\em dynamical symmetry}. Each dynamical symmetry
$\tilde{h}$ induces a map $h : \Gamma \mapsto \Gamma$ that preserves the
symplectic structure $\Omega$, so $h$ is a symplectomorphism of the physical
phase space $(\Gamma,\Omega)$.

An infinitesimal extended dynamical symmetry $\overline{dh}$ is generated via
Poisson brackets by a function $-\bar{H}dt : {\mathcal P} \mapsto {\mathcal
  P}$ (the sign is chosen for later convenience). The restriction $\tilde{H}$
of $\bar{H}$ to ${\mathcal C}$ is a Dirac observable and so it defines a
unique function $H$ on the physical phase space $\Gamma$. The function $-Hdt$
generates the infinitesimal symplectomorphism $dh$ in $\Gamma$ that is induced
by $\overline{dh}$, for proofs, see \cite{B-O}, \cite{H1} and \cite{H2}. (We
denote infinitesimal maps by a symbol composed from ``$d$'' and a letter; the
letter itself has no further meaning.)

The infinitesimal transformations generated by $({\mathcal H}[\vec{N}] +
{\mathcal H}_\infty[\vec{N}_\infty])dt$ are dynamical symmetries for all
$\vec{N}$ satisfying PGAC. Their Poisson brackets at the constraint surface
form the Lie algebra of Poincar\'{e} group \cite{B-O}. They have in general
non-vanishing Poisson brackets with other Dirac observables generating a
non-trivial dynamics for them. We shall give an interpretation of this
dynamics in the next two subsections.

\subsection{Transversal surfaces}
\label{sec:trans}
A transversal surface $\mathcal T$ is a submanifold of $\mathcal C$ that
intersects each c-orbit exactly once and transversally; $\mathcal T$ is a
section of the quotient ${\mathcal C}/\lambda$. Each transversal surface
inherits a two-form $\Omega_{\mathcal T}$ from $\mathcal C$ by pull back of
$\tilde{\Omega}$ to $\mathcal T$ by the injection map of $\mathcal T$ into
$\mathcal C$. It is not only closed but also non-degenerate, so $({\mathcal
  T}, \tilde{\Omega})$ is a symplectic manifold (see also \cite{B-O}, where
transversal surface is called ``gauge condition'').

The projection $\pi_{\text{phs}}$ maps $\mathcal T$ to $\Gamma$ and its
restriction $\pi_{\text{phs}}\vert_{\mathcal T}$ is actually a bijection
having an inverse. The map $\pi_{\text{phs}}\vert_{\mathcal T}$ can be shown
to be a symplectomorphism of the spaces $({\mathcal T}, \Omega_{\mathcal T})$
and $(\Gamma,\Omega)$. Thus, each transversal surface is a ``model'' of the
physical phase space. If we wish to calculate $\Omega$ in the coordinates
$\{o^i\}$, then we just have to calculate $\Omega_{\mathcal T}$ in the
coordinates $\{\tilde{o}^i\}$.

An important issue is the relation between a transversal surface $\mathcal T$
and Cauchy surfaces in solution space-times. Such a relation is provided by a
gauge fixing $\sigma$. As it has been shown above, $\sigma$ can be represented
as a set of functions $\Psi(o,X)$, where $\{o^i\}$ are coordinates on $\Gamma$
and $X^\mu$ coordinates on $\mathcal M$. Consider the spacetime
$\bigl({\mathcal M},\Psi(o,X)\bigr)$; a Cauchy surface in such a spacetime can
be described by an embedding $\hat{X}(o) : \Sigma \mapsto {\mathcal M}$ given
by the set of functions $X^\mu(o,x)$. Having these functions and the solution
$\Psi(o,X)$, we can calculate the corresponding Cauchy datum
$(\Psi_{\hat{X}(o)},\Pi_{\hat{X}(o)})$ and so determine a point at $\mathcal
C$. More specifically, this point lies at the c-orbit $\lambda(o)$
corresponding to the point $o \in \Gamma$: $\lambda(o) =
\pi_{\text{phs}}^{-1}o$. In this way, there is a unique point
$(\Psi_{\hat{X}(o)},\Pi_{\hat{X}(o)})$ at $\mathcal C$ for each $\lambda$, and
this defines a transversal surface $\mathcal T$.

Let us denote the resulting map that sends the set of embeddings
$\{\hat{X}(o)\}$ to the transversal surface $\mathcal T$ by $\chi_\sigma$:
\[
  \chi_\sigma\{\hat{X}(o)\} = {\mathcal T}\ .
\]
The functions $\Psi(o,X)$ and $X(o,x)$ must be sufficiently smooth or else
$\mathcal T$ would be rather jumpy. 

One can show that $\chi_\sigma$ can even be a bijection between the space of
embedding sets $\{X(o,x)\}$ and the space of transversal surfaces $\mathcal
T$. The condition is that the solutions $\Psi(o,X)$ do not admit any spacetime
symmetry \cite{H-Kij}. Then each transversal surface and $\sigma$ determine
together a set of embeddings $\{X(o,x)\}$, one embedding for each solution
spacetime $\bigl({\mathcal M},\Psi(o,X)\bigr)$:
\[
  \{X(o,x)\} = \chi_\sigma^{-1}{\mathcal T}\ .
\]
One can say that, given a gauge, each transversal surface defines a {\em
  many-finger-time} instant in each solution spacetime. Transversal surfaces
can, therefore, also be called many-finger-time levels.

An interesting point is that the variables $o^i$ and
$\{X(o,x)\}$---coordinates on the physical phase space and a set of
embeddings, one for each solution---form a coordinate system on $\mathcal C$
if a gauge $\sigma$ is specified. They are called generalized\footnote{The
  embeddings introduced originally by Kucha\v{r} \cite{embed1}, \cite{embed2}
  were $o$-independent.} {\em Kucha\v{r} variables}, and they constitute a
useful tool for many calculations because they provide a neat division of
variables into gauge, physical and dependent degrees of freedom on one hand
and a bridge between the four-dimensional and the three-dimensional pictures
on the other.

\subsection{Time evolution}
Dirac observables are constant along whole solutions and so they are not only
gauge invariant but a kind of integrals of motion. It seems that their
dynamics is trivial: they just stay constant. Bergmann \cite{berg}
characterized that as ``frozen dynamics''.

However, as early as 1949, Dirac \cite{D1} has put forwards a theory of time
evolution that is based on transversal surfaces and symmetries and that leads
to a non trivial time evolution of Dirac observables. The idea is that a
physically sensible evolution results if the ``pure'' dynamics (such as
staying constant) is compared with a fiducial ``zero'' dynamics defined by a
symmetry. The account presented here is a generalization \cite{H1}, \cite{H2}
of Dirac theory. It can also be understood as an illustration of the discovery
\cite{K-T} that the Hamiltonian dynamics needs a frame of reference and that
different frames lead to differently looking time evolutions of one and the
same system, see also \cite{C-J-K}.

Let $\tilde{h}$ be dynamical symmetry and $\mathcal T$ a transversal
surface. Then $\tilde{h}{\mathcal T} = {\mathcal T}'$ is another transversal
surface and $\tilde{h}$ is a symplectomorphism between the spaces $({\mathcal
  T},\Omega_{\mathcal T})$ and $({\mathcal T}',\Omega_{{\mathcal T}'})$. Thus,
$\tilde{h}$ can shift many-finger-time levels.

Suppose a one-dimensional group of extended dynamical symmetries $\bar{h}(t)$
is given; $t$ is the parameter of the group, $\bar{h}(t_1) \cdot \bar{h}(t_2)
= \bar{h}(t_1 + t_2)$. Let the generator of the group be $-\bar{H}$, and let
the induced groups and generators on $\mathcal C$ and $\Gamma$ be denoted by
$\tilde{h}(t)$, $h(t)$ and $-H$. We use the group $\tilde{h}(t)$ to build up a
reference system in $\mathcal C$ with respect to which we shall describe the
motion represented by c-orbits in a similar way as the particle world lines in
Minkowski spacetime are described by their coordinates in an inertial frame.

Let $\mathcal T$ be a transversal surface. Define ${\mathcal T}_t :=
\tilde{h}(t){\mathcal T}_0$. These surfaces form a one-dimensional family that
we denote by $\{{\mathcal T}_t\}$. The family determines a subset
$\eta_\lambda(t)$ of any c-orbit $\lambda$ by
\[
  \eta_\lambda(t) := {\mathcal T}_t \cap \lambda\ .
\]
The curve $\eta_\lambda(t)$ lies in $\lambda$ and is one-dimensional in
spite of $\lambda$ being itself many (or even infinitely many) dimensional. We
call $\eta_\lambda(t)$ a trajectory of the system with respect to the family
$\{{\mathcal T}_t\}$.

Let $\tilde{o}_0$ be a Dirac observable representing some measurement done at
the time level\footnote{Indeed, a given Dirac observable cannot, in general,
  be measured anywhere on the constraint surface but only in a neighbourhood
  of a transversal surface that belongs to the definition of the observable,
  see \cite{H2}.} ${\mathcal T}_0$. The values of the function
$\tilde{o}_0\vert_{{\mathcal T}_0}$ on ${\mathcal T}_0$ give results of
measurement at each point of ${\mathcal T}_0$, that is, at each instantaneous
state of the system.

We define {\em the same measurement} at the time ${\mathcal T}_t$ to be
represented by Dirac observable $\tilde{o}_t := \tilde{o}_0 \circ
\tilde{h}(-t)$. The function $\tilde{o}_t$ is the image of $\tilde{o}_0$ by
the map $\tilde{h}(t)$: it gives the same results at the states that are
related by the symmetry:
\[
  \tilde{o}_t\bigl(\tilde{h}(t)p\bigr) = \tilde{o}_0(p)
\]
for any point $p \in {\mathcal C}$. Actually, the set $\{\tilde{o}_t|t \in
{\mathbb R}\}$ of Dirac observables is a special case of Rovelli's ``evolving
constants of motion'' \cite{Revolv}.

Now, we are ready to define the time evolution.
\begin{df}Time evolution is the change in the results of the same measurement
  done at different times along a dynamical trajectory of the system.
\end{df}

Clearly, these results are given by the function $\tilde{o}_t
\bigl(\eta_\lambda(t)\bigr)$ for the measurement represented by
$\{\tilde{o}_t|t \in \mathbb R\}$ and the trajectory $\lambda$. The function
can be written in two ways.

\paragraph{Schr\"{o}dinger picture} By substituting from the definition of
$\tilde{o}_t$, we have
\[
  \tilde{o}_t\bigl(\eta_\lambda(t)\bigr) =
  \tilde{o}_0\bigl(\tilde{h}(-t)\eta_\lambda(t)\bigr)\ .
\]
Let us define
\[
  \tilde{\xi}_\lambda(t) := \tilde{h}(-t)\eta_\lambda(t)\ ;
\]
$\tilde{\xi}_\lambda(t)$ is a curve in ${\mathcal T}_0$ and can be viewed as a
``time-dependent state of the system.'' Then we can write
\begin{equation}
  \tilde{o}_t\bigl(\eta_\lambda(t)\bigr) = \tilde{o}_0\vert_{{\mathcal T}_0}
  \bigl(\tilde{\xi}_\lambda(t)\bigr)\ .
\label{71.1}
\end{equation}
The measurement is represented by a single (time-independent) Schr\"{o}dinger
observable $\tilde{o}_0\vert_{{\mathcal T}_0}$.

\paragraph{Heisenberg picture} Since $\tilde{o}_t$ is a Dirac observable for
each $t$, it is constant along c-orbits, and we find
\[
  \tilde{o}_t\bigl(\eta_\lambda(t)\bigr) =
  \tilde{o}_t\bigl(\eta_\lambda(0)\bigr)\ .
\]
All quantities can again be taken at ${\mathcal T}_0$:
\begin{equation}
  \tilde{o}_t\bigl(\eta_\lambda(t)\bigr) = \tilde{o}_t\vert_{{\mathcal
  T}_0}\bigl(\eta_\lambda(0)\bigr)\ .
\label{72.1}
\end{equation}
Now, the state is described by a single point $\eta_\lambda(0)$ at ${\mathcal
  T}_0$ (time independence) while the measurement is described by a set of
functions $\tilde{o}_t\vert_{{\mathcal T}_0}$ on ${\mathcal T}_0$,
constituting a time-dependent Heisenberg observable
$\{\tilde{o}_t\vert_{{\mathcal T}_0}|t \in {\mathbb R}\}$.

The dynamical equation for the Schr\"{o}dinger state $\tilde{\xi}(t)$ or the
Heisenberg observable $\{\tilde{o}_t\vert_{{\mathcal T}_0}|t \in {\mathbb
  R}\}$ can be calculated with the following results: The Schr\"{o}dinger
state $\tilde{\xi}(t)$ is an integral curve in $({\mathcal
  T}_0,\Omega_{{\mathcal T}_0})$ of the Hamiltonian vector field
$d\tilde{H}\vert_{{\mathcal T}_0}^{\#}$ of $\tilde{H}\vert_{{\mathcal T}_0}$:
\begin{equation}
  \frac{d\tilde{\xi}_\lambda(t)}{dt} = d\tilde{H}\vert_{{\mathcal T}_0}^{\#}\ .
\label{73.1}
\end{equation}
The Heisenberg observable $\{\tilde{o}_t\vert_{{\mathcal T}_0}\}$ must satisfy
the differential equation
\begin{equation}
  \frac{d\tilde{o}_t\vert_{{\mathcal T}_0}}{dt} =
  \{d\tilde{o}_t\vert_{{\mathcal T}_0}, \tilde{H}\vert_{{\mathcal
  T}_0}\}_{\Omega\vert_{{\mathcal T}_0}}\ ,
\label{73.2}
\end{equation}
where $\{\cdot,\cdot\}_{\Omega\vert_{{\mathcal T}_0}}$ is the Poisson bracket
of $({\mathcal T}_0,\Omega_{{\mathcal T}_0})$.

All definitions and equations hitherto written down depend on the choice of
${\mathcal T}_0$. This leads to the impression that our dynamics depends not
only on the chosen symmetry group $\tilde{h}(t)$ but also on ${\mathcal T}_0$.
This impression may be strengthened if one reads Dirac's paper. However,
Dirac's choice of the group was related to his choice of the surface. Here,
these two choices are independent.

Actually, the exposition given as yet has aimed at some physical
interpretation of the final result, which comes only now: The
symplectomorphism $\pi_{\text{phs}}\vert_{{\mathcal T}_0}$ maps the
Eqs.~(\ref{73.1}) and (\ref{73.2}) at ${\mathcal T}_0$ to the following
equations at $\Gamma$ 
\begin{equation}
  \frac{d\xi_\lambda}{dt} = dH^{\#}
\label{74.1}
\end{equation}
and 
\begin{equation}
  \frac{do_t}{dt} = \{o_t,H\}_\Omega\ ,
\label{74.2}
\end{equation}
where
\[
  \xi_\lambda(t) := \pi_{\text{phs}}\vert_{{\mathcal
  T}_0}\tilde{\xi}_\lambda(t)\ ,\quad \tilde{o}_t\vert_{{\mathcal T}_0} = o_t
  \circ \pi_{\text{phs}}\vert_{{\mathcal T}_0}\ .
\]
Eqs.~(\ref{74.1}) and (\ref{74.2}) have the same form for any choice of
${\mathcal T}_0$ and depend only on the group $h(t)$.

The symmetry group $h(t)$ is to be chosen as a suitable subgroup of the
asymptotic symmetries. To be sure, asymptotic symmetries cannot be extended to
the whole of $\mathcal M$ in a unique way (a gauge-dependent extension has
been constructed at the end of subsection \ref{sec:sol}). However, the action
of such extended map on a gauge-invariant quantity does not depend on the way
it has been extended. The dynamics constructed in this way is, therefore,
gauge invariant and it seems to be physically sound.

\section{A model: gravitating shell}
\label{sec:model}
This section will illustrate the notions and methods introduced in the
preceding one. At the same time, the dynamics of the shell will be given a
form that will be suitable for quantization.

\subsection{Space of solutions, gauge group \\ and asymptotic symmetries}
\label{sec:mspace}
The model consists of a spherically symmetric thin shell of light-like matter
surrounded by its own gravitational field. Hence, there are two geometrical
objects in $\bar{\Psi}$: a metric $g_{\mu\nu}$ describing the gravitational
field and a three-dimensional light-like surface $\bar{S}$ that is a
trajectory of the shell. The background manifold $\bar{\mathcal M}$ is
${\mathbb R}^3 \times {\mathbb R}$, where ${\mathbb R}^3$ is the manifold of
Cauchy surface. The dynamical equations are Einstein's and matter
equations. Any solution $\bar{\Psi}$ can be constructed by sticking together a
piece of Schwarzschild solution of mass parameter $M$  with the meaning of
gravitational radius, and a piece of Minkowski spacetime, along a
spherically symmetric null hypersurface so that the points with the same
radius coordinate cover each other (see, e.g., \cite{H-Kie}).

The spherical symmetry enables us to reduce the dimension of the problem by
two. The effective spacetime can be considered as the space of the rotation
group orbits; the background manifold $\mathcal M$ is ${\mathbb R} \times
{\mathbb R}_+$, where ${\mathbb R}_+ := (0,\infty)$. The four-dimensional
metric decomposes into a two-dimensional metric $g_{AB}$ on $\mathcal M$ and a
scalar field $R$ on $\mathcal M$ (radius of the rotation group orbit). The
shell trajectory becomes a curve $S$. These are the three geometrical objects
in $\Psi$.

Luckily, for this simple model, all solutions can be explicitly written down.
Thus, we can make a list of all physically different solutions to a basis of
our analysis. The list can best be given in a particular gauge determined by a
covariant gauge fixing $\sigma$. To specify $\sigma$, we choose the
coordinates $U$ and $V$ on $\mathcal M$ with ranges $U \in {\mathbb R}$ and $V
\in (U,\infty)$. The boundary of $\mathcal M$ (that does not belong to it)
consists of three pieces: $\partial_0{\mathcal M}$ given by the equation $V =
U$, with a coordinate $T_0 := (U + V)/2$, $T_0 \in {\mathbb R}$,
$\partial_+{\mathcal M}$ defined by the limit $V \rightarrow \infty$ and
described by coordinate $U_\infty = U$, $U_\infty \in {\mathbb R}$, and
$\partial_-{\mathcal M}$ defined by the limit $U \rightarrow -\infty$ and
described by coordinate $V_\infty = V$, $V_\infty \in {\mathbb R}$. With
respect to the fixed coordinates $U$ and $V$, we set conditions on the
components of the representative metric, scalar field and shell trajectory as
follows (for more detail and motivation, see \cite{H-Kie} and \cite{H-Kou}).
\begin{enumerate}
\item The coordinates $U$ and $V$ are double null coordinates:
\[
 ds^2 = -A(U,V)dUdV\ .
\]
\item For out-going shells, $S$ is defined by $U = $ const, and the coordinate
  $U$ coincides with a retarded time at $\partial_+{\mathcal M}$.
\item For in-going shells, $S$ is defined by $V = $ const, and the coordinate
  $V$ coincides with an advanced time at $\partial_-{\mathcal M}$.
\item The functions $A(U,V)$ and $R(U,V)$ are continuous at the shell.
\end{enumerate}
Then the representative solutions can be described as follows (see
\cite{H-Kie} and \cite{H-Kou}). We introduce a parameter $\eta$ with two
values $+1$ and $-1$ to define the direction of radial motion of the shell:
expanding for $\eta = +1$ and contracting for $\eta = -1$.

\paragraph{Out-going shell} ($\eta = +1$).
\begin{enumerate}
\item The shell trajectory is given by
\[
  U = w\ ,
\]
$w \in {\mathbb R}$ being a measurable parameter: the retarded arrival time of
the out-going shell.
\item Left from the shell, $U > w$, the two-metric $A(U,V)$ and the scalar
  field $R(U,V)$ are given by
\begin{equation}
  A = 1\ ,\quad R = \frac{-U+V}{2}\ .
\label{80.2}
\end{equation}
\item Right from the shell, $U < w$, they are given by
\begin{equation}
  A = \frac{1}{\kappa(f_+)e^{\kappa(f_+)}}\frac{-w+V}{4M}
  \exp\left(\frac{-U+V}{4M}\right)\ ,
\label{80.3} 
\end{equation}
and 
\begin{equation}
  R = 2M\kappa(f_+)\ ,
\label{80.4}
\end{equation}
where
\begin{equation}
  f_+ = \left(\frac{-w+V}{4M} - 1\right)\exp\left(\frac{-U+V}{4M}\right)\ .
\label{80.5}
\end{equation}
\end{enumerate}
The Kruskal function $\kappa$ is defined by its inverse
\[
  \kappa^{-1}(x) = (x-1)e^x
\]
on the interval $x \in (0,\infty)$. 

At the shell,
\[
  \lim_{U\rightarrow w_-}f_+ = \kappa^{-1}\left(\frac{-w+V}{4M}\right)\ ,
\]
hence, we have $R \rightarrow (-w+V)/2$ and $A \rightarrow 1$ and the metric
is continuous, but its derivatives have a jump. In this way, these coordinates
carry the differential structure $C^1$ that is determined by the metric. The
stress-energy tensor of the shell matter can be calculated by the formula
given in \cite{B-I}. Let us denote this solution by $\Psi_+(M,w;U,V)$.

The boundary $R \rightarrow 0$ is given by $V \rightarrow U$ for $U > w$ and
by $f_+ \rightarrow -1$ or 
\begin{equation}
  V \rightarrow w + 4M\kappa\left[-\exp\left(\frac{U - w}{4M}\right)\right]
\label{81.1}
\end{equation}
for $U < w$. The curve defined by the right-hand side of (\ref{81.1}) is
space-like running from the point $(U,V) = (w,w)$ to the point $(U,V) =
(-\infty,w + 4M)$. It is the boundary of the open domain ${\mathcal D}_\Psi$
in $\mathcal M$ where the representative solution $({\mathcal M},\Psi_+)$ is
well defined. The curve $V = w + 4M$ is the horizon, $R = 2M$.

\paragraph{In-going shell} ($\eta = -1$).
\begin{enumerate}
\item The shell trajectory is given by
\[
  V = w\ ;
\]
$w \in {\mathbb R}$ is the advanced departure time of the in-going shell.
\item Left from the shell, $V < w$, the two-metric $A(U,V)$and the scalar
  field $R(U,V)$ are given by
\begin{equation}
  A = 1\ ,\quad R = \frac{-U+V}{2}\ .
\label{80.2-}
\end{equation}
\item Right from the shell, $V > w$, they are given by
\begin{equation}
  A = \frac{1}{\kappa(f_-)e^{\kappa(f_-)}}\frac{-U+w}{4M}
  \exp\left(\frac{-U+V}{4M}\right)\ ,
\label{80.3-} 
\end{equation}
and 
\begin{equation}
  R = 2M\kappa(f_-)\ ,
\label{80.4-}
\end{equation}
where
\begin{equation}
  f_- = \left(\frac{-U+w}{4M} - 1\right)\exp\left(\frac{-U+V}{4M}\right)\ .
\label{80.5-}
\end{equation}
\end{enumerate}
Let us denote this solution by $\Psi_-(M,w;U,V)$. Again, the metric is
continuous at the shell. The boundary of the domain ${\mathcal D}_\Psi$ is
given by the equation $f_- = -1$. Observe that the solutions $\Psi_-(M,w;U,V)$
fill in the lower corner of $\mathcal M$ and so every point of the manifold
$\mathcal M$ is used by both solutions $\Psi_+(M,w;U,V)$ and $\Psi_-(M,w;U,V)$.

The first employment of our list is the specification of the physical phase
space. It has two components, $\Gamma_+$ and $\Gamma_-$ corresponding to $\eta
= +1$ and $\eta = -1$. They have both the manifold structure of ${\mathbb R}_+
\times {\mathbb R}$, where $M \in {\mathbb R}_+$ and $w \in {\mathbb R}$. $M$
and $w$ can serve as coordinates on $\Gamma_\eta$. It is not yet clear that
these coordinates are regular; this ought to follow from the regularity of the
symplectic form of the physical phase space written with respect of these
coordinates. The symplectic form cannot be inferred from our list of solutions
and will be calculated later.

The group of general four-dimensional diffeomorphisms can also be reduced by
the symmetry: we admit only diffeomorphisms that commute with rotations so
that the rotation group orbits are mapped onto such orbits. Clearly, regular
centre points are sent only to regular centre points---so that the regular
centre is invariant---while a spherical orbit can be sent to any spherical
orbit. Hence, the reduced group is Diff$\mathcal M$ for our two-dimensional
background manifold $\mathcal M$.

The second employment of the solution list given above is to find 
symmetries of the system. The symmetries will help us to decompose the group
Diff$\mathcal M$, to define the most important observables and to construct
the dynamics. To find the symmetries, we observe that some pairs of our
representative solutions are isometric or conformally related to each
other. For each such pair, there is a unique element of Diff$\mathcal M$ that
implements the relation.

\paragraph{Time shifts} Clearly, the solutions $\Psi_\eta(M,w;U,V)$ and
$\Psi_\eta(M,w+t;U,V)$ are isometric for any $t \in {\mathbb R}$, $M$, $w$ and
$\eta$. The corresponding diffeomorphism $\varphi_H(t)$ is given by 
\begin{equation}
  \varphi_H(t) : (U,V) \mapsto (U + t, V + t)\ 
\label{fiH}
\end{equation}
For example, if $\eta = +1$, the shell is sent to $U = w + t$, the new metric and the scalar are, for
$U > w+t$,
\[
  A = 1\ ,\quad R = \frac{-U+V}{2}\ ,
\]
while, for $U < w+t$, 
\[
  A = \frac{1}{\kappa(f'_+)e^{\kappa(f'_+)}}\frac{-w-t + V}{4M}
  \exp\left(\frac{-U+V}{4M}\right)\ , 
\]
and
\[
  R = 2M\kappa(f'_+)\ ,
\]
where
\[
  f'_+ = \left(\frac{-w-t+V}{4M} - 1\right)\exp\left(\frac{-U+V}{4M}\right)\ .
\]
Hence, these are just the fields (\ref{80.2})--(\ref{80.5}) with $w \mapsto w
+ t$. The proof for $\eta =-1$ is analogous. Observe that the map
$\varphi_H(t)$ is well defined everywhere on $\mathcal M$ and independent of
the parameters $\eta$, $M$ and $w$. Hence, all such maps form a
one-dimensional group $G_H$ with parameter $t$. The group acts on the physical
phase space: $\varphi_H(t) : (M,w) \mapsto (M,w+t)$. Finally, its action can
be uniquely extended to the boundary:
\[
  U_\infty \mapsto U_\infty + t\ ,\quad V_\infty \mapsto V_\infty + t\ ,\quad
  T_0 \mapsto T_0 + t\ .
\]

\paragraph{Dilatations} Clearly, the solutions $\Psi_\eta(M,w;U,V)$ and
$\Psi_\eta(e^sM,e^sw;U,V)$ are conformally related for all $s \in {\mathbb
  R}$, $\eta$, $M$ and $w$. On the other hand, they are sent into each other
by a composite map $\Bigl(\varphi_D(s),\omega\bigl(\varphi_D(s)\bigr)\Bigr)$,
where
\begin{equation}
  \varphi_D(s) : (U,V) \mapsto (e^sU,e^sV)
\label{fiD}
\end{equation}
is a diffeomorphism and 
\[
  \omega\bigl(\varphi_D(s)\bigr) : g_{\mu\nu} \mapsto e^{2s}g_{\mu\nu}
\]
is a constant conformal deformation (CCD) that acts only on the
(four-dimensional) metric: it scales the two-dimensional metric $g_{AB}$ by
$e^{2s}$ and the scalar field $R$ by $e^s$.

Actually, the dilatation is nothing but a rescaling: all quantities with a
dimension are rescaled by the corresponding power of $e^s$. The coordinates
$U$ and $V$ have acquired their scale from the metric through the gauge
fixing. The Newton constant G must also be rescaled: G $\mapsto e^{2s}$G!

To see that the map (\ref{fiD}) makes the job, let us restrict ourselves to
the case $\eta = +1$ and study the action of the map on the objects 
in the spacetime defined by Eqs.~(\ref{80.2})--(\ref{80.5}). The metric is
transformed by $\varphi_D(s)_*$ to
\[
  ds^2 = - e^{-2s}A(e^{-s}U,e^{-s}V)dUdV\ ,
\]
the scalar field to
\[
 R = R(e^{-s}U,e^{-s}V)
\]
and the shell position changes as follows
\[
   w \mapsto e^s w\ .
\]
Thus, for $U > e^s w$, the new metric $A$ is $e^{-2s}$ and the new scalar $R$
is
\[
  e^{-s}\frac{-U+V}{2}\ ;
\]
for $U < e^s w$, the new metric $A$ is 
\[
  A = \frac{e^{-2s}}{\kappa(f'_+)e^{\kappa(f'_+)}}\frac{-we^s +
  V}{4Me^s} \exp\left(\frac{-U+V}{4Me^s}\right)
\]
and the scalar $R$,
\[
 R = 2M\kappa(f'_+)\ ,
\]
where
\[
  f'_+ = \left(\frac{-we^s + V}{4Me^s} - 1\right)
  \exp\left(\frac{-U+V}{4Me^s}\right)\ .
\]
The subsequent CCD by $e^{2s}$ brings $A$ and $R$ to
\begin{equation}
  A = 1\ ,\quad R = \frac{-U+V}{2}
\label{90.1}
\end{equation}
for $U > we^s$, to 
\begin{equation}
  A = \frac{1}{\kappa(f'_+)e^{\kappa(f'_+)}}\frac{-we^s +
  V}{4Me^s} \exp\left(\frac{-U+V}{4Me^s}\right)
\label{90.2} 
\end{equation}
and
\begin{equation}
  R = 2Me^s\kappa(f'_+)\ ,
\label{90.3}
\end{equation}
for $U < we^s$, and the shell stays at $U = we^s$. But the fields
(\ref{90.1})--(\ref{90.3}) are just the same as (\ref{80.2})--(\ref{80.5})
except for $M$ being changed to $Me^s$ and $w$ to $we^s$.

We observe that the map $\Bigl(\varphi_D(s),
\omega\bigl(\varphi_D(s)\bigr)\Bigr)$ is well defined on the whole $\mathcal
M$ and independent of $\eta$, $M$ and $w$. Since each CCD commutes with any
diffeomorphism, the dilatations form a one-dimensional group $G_D$ with
parameter $s$. The part $\varphi_D(s)$ form themselves a group $G_{D\varphi}$
that is a subgroup of Diff$\mathcal M$. The dilatations act on the physical
phase space:
\[
  \Bigl(\varphi_D(s),\omega\bigl(\varphi_D(s)\bigr)\Bigr) : (\eta,M,w) \mapsto
  (\eta,e^sM,e^sw)\ .
\]
Finally, dilatations act on the boundary
\[
  U_\infty \mapsto e^sU_\infty\ ,\quad V_\infty \mapsto e^sV_\infty\ ,\quad
  T_0 \mapsto e^sT_0\ . 
\]

The time shifts and dilatations generate a group. Each element of the group
can be obtained in a unique way as a dilatation followed by a time shift. Let
us write it in the form $\Bigl(\varphi_H(t), \varphi_D(s),
\omega\bigl(\varphi_D(s)\bigr)\Bigr)$. If the action of the constituent maps
on, say, the parameter $w$ is taken into account, the following composition
law can be found
\begin{multline*}
  \Bigl(\varphi_H(t_1),\varphi_D(s_1),\omega\bigl(\varphi_D(s_1)\bigr)\Bigr)
  \times \\
  \Bigl(\varphi_H(t_2),\varphi_D(s_2),\omega\bigl(\varphi_D(s_2)\bigr)\Bigr) 
  = \\ \Bigl(\varphi_H(t_1+e^{s_1}t_2), \varphi_D(s_1+s_2),
  \omega\bigl(\varphi_D(s_1+s_2)\bigr)\Bigr)\ . 
\end{multline*}
It shows that the group has the structure of semi-direct product of the
multiplicative ${\mathbb R}_+$ with the additive $\mathbb R$ abelian groups
($t \in {\mathbb R}$, $e^s \in {\mathbb R}_+$). This group is usually called
{\em affine group} $\mathcal A$ on $\mathbb R$. The time shifts form a normal
subgroup of $\mathcal A$: they play the role of the abelian factor in the
semi-direct product.

\paragraph{Time reversal} The two solutions $\Psi_\eta(M,w;U,V)$ and
$\Psi_{-\eta}(M,-w;U,V)$ are isometric. The isometry is implemented by the
diffeomorphism $\varphi_I$ defined by
\[
  \varphi_I : (U,V) \mapsto (-V,-U)\ .
\]
It is well-defined everywhere in $\mathcal M$ and independent of $\eta$, $M$
and $w$. Its action on the boundary is
\[
  U_\infty \mapsto -V_\infty\ ,\quad V_\infty \mapsto -U_\infty\ ,\quad T_0
  \mapsto -T_0\ ,
\]
and that on the physical phase space is
\[
  I : (\eta,M,w) \mapsto (-\eta,M,-w)\ .
\]
 
The full group of symmetries generated by all three kinds of maps is denoted
by $G_\sigma$. The index $\sigma$ is to remind that the action of the group on
$\mathcal M$ has been obtained with the help of our solution list that
represents the gauge fixing $\sigma$; a different gauge fixing would lead to
different action on $\mathcal M$. However, the action of $G_\sigma$ on the
physical phase space as well as some aspects of its action on the boundary are
gauge independent. So is the group structure of $G_\sigma$, which is that of
the affine group ${\mathcal A}_I$ with inversion.

The action of $G_\sigma$ on the boundary $\partial{\mathcal M}$ defines a
group of boundary transformations that will be denoted by $G_b$. The group
$G_b$ consists of pure diffeomorphisms (not polluted by any CCD's). Each
element of $G_\sigma$ determines a unique element of $G_b$ and this map
between $G_\sigma$ and $G_b$ is an isomorphism because $G_\sigma$ acts truly
on the boundary.

There is one point about the structure of $G_b$ that will be useful later: the
subgroup generated by time shifts and the time reversal is invariant (normal).
Hence, there is a homeomorphism $\pi_D$ of $G_b$ onto the factor group, which
is $G_{D\varphi}$, $\pi_D : G_b \mapsto G_{D\varphi}$. This means that the
amount of dilatation hidden in any element of $G_b$ is well-defined.

\paragraph{Decomposition of Diff$\mathcal M$} Let us choose a subgroup $G$ of
Diff$\mathcal M$ according to the following rules. The diffeomorphism
$\varphi$ is an element of $G$ if
\begin{enumerate}
\item the map $\varphi$ has a differentiable extension to the boundary of
  $\mathcal M$. Hence, it defines a map $\varphi_b : \partial{\mathcal M}
  \mapsto \partial{\mathcal M}$,
\item The action of $\varphi$ on the boundary coincides with that of an element
  of the symmetry group $G_\sigma$, i.e., $\varphi_b \in G_b$.
\end{enumerate}
The group $G$ is a proper subgroup of Diff$\mathcal M$. It is the only part of
Diff${\mathcal M}$ that is interesting for the physics of the system. It is
easy to check that the map $\pi_b : G \mapsto G_b$, the existence of which
follows from the above definition, is a homeomorphism (preserves
multiplication). The kernel $G_0$ of $\pi_b$ is a set of elements of $G$ that
is mapped by $\pi_b$ to the identity of $G_b$. Such a kernel must be a normal
subgroup of $G$. The group $G_0$ is the {\em gauge group} of the system.

The group $G$ itself needs a correction by CCD's at those elements that can be
said to contain dilatation. The amount of dilatation in $\varphi \in G$ is
obviously given by $\pi_D\pi_b\varphi$. Hence, the pairs $\bigl(\varphi,
\omega(\pi_D\pi_b\varphi)\bigr)$ are {\em symmetries} of our system. They form
a group with the same structure as $G$ because CCD's commute with
diffeomorphisms. Let us denote the group by $G_c$ ($c$ for corrected). 

The action of the symmetry group $G_c$ on the boundary has a gauge-invariant
aspect. This can be revealed after attaching the scri to the solutions.
Consider, for example, the spacetime $\bigl({\mathcal M},
\Psi_+(M,w;U,V)\bigr)$. The ${\mathcal I}^+(+1,M,w)$ of this spacetime is
one-dimensional and a coordinate $\tilde{U}$ can be chosen along it so that
the difference $\tilde{U}_2 - \tilde{U}_1$ gives the interval of retarded time
along ${\mathcal I}^+(+1,M,w)$. In the spherically symmetric case, such an
interval is uniquely determined. However, $\tilde{U}$ clearly contains more
information than just that about intervals: an origin of $\tilde{U}$ has also
been chosen. This origin is the only non-trivial remainder, in the spherically
symmetric case, of what can be called {\em asymptotic frame}. In an analogous
way, an inertial coordinate system in Minkowski spacetime contains not only
information about Minkowski interval (metric), but also that on the underlying
inertial frame.

Let us attach ${\mathcal I}^+(+1,M,w)$ so that
\begin{equation}
  \tilde{U} = U_\infty\ .
\label{scri+}
\end{equation}
This is, in any case, possible because $U_\infty$ {\em is} a retarded time
along $\partial_+{\mathcal M}$ for the spacetime $\bigl({\mathcal M},
\Psi_+(M,w;U,V)\bigr)$ so that the intervals match. However, Eq.~(\ref{scri+})
represents also a judicious choice of origins, that is, asymptotic frames, for
each solution $\Psi_+(M,w;U,V)$. The idea behind the choice is that $w$
keeps the role of the retarded arrival time of the out-going shells also with
respect to ${\mathcal I}^+(+1,M,w)$. We shall see later how this choice of
frame of asymptotic reference influences the form of the asymptotic action of
the symmetry group and the calculation of the symplectic form.

At $\partial_-{\mathcal M}$, where $U \rightarrow -\infty$
and $V \in (u + 4M,\infty)$, we must recover the Schwarz\-schild advanced time
$\tilde{V}$ as a function of $V$ in order to attach ${\mathcal
  I}^-(+1,M,w)$. An easy calculation yields:
\begin{equation}
  \tilde{V} = V_\infty + 4M\ln\left(\frac{-w+V_\infty}{4M} - 1\right)\ ,\quad
  V_\infty  = w + 4M\kappa\left[\exp\left(\frac{\tilde{V}}{4M}\right)\right]\
  . 
\label{84.1}
\end{equation}
As $V \in (w + 4M,\infty)$, $\tilde{V}$ runs through $\mathbb R$. The advanced
time
interval is just $\tilde{V}_2 - \tilde{V}_1$. Eq.~(\ref{84.1}) replaces
(\ref{scri+}) and determines the attachment of ${\mathcal I}^-(+1,M,w)$ to the
spacetime $\bigl({\mathcal M}, \Psi_+(M,w;U,V)\bigr)$. Observe that ${\mathcal
  I}^-(+1,M,w)$ takes only a part of $\partial_-{\mathcal M}$. 

Finally, ${\mathcal I}^0(+1,M,w)$ is a ``stretched'' space-like infinity
$i^0$: it has the structure of $\mathbb R$ and the coordinate $T_\infty$ on it
can be defined as limit point of a constant Schwarzschild time $T = T_\infty$
surfaces. Such surfaces are defined in the coordinates $U$ and $V$ by
\[
  \frac{U + \tilde{V}(V)}{2} = T_\infty\ .
\]
We obtain from Eq.~(\ref{84.1}):
\[
  \frac{U}{2} + \frac{V}{2} + 2M\ln\left(\frac{-w + V}{4M} - 1\right) =
  \text{const} 
\]
as $U \rightarrow -\infty$ and $V \rightarrow \infty$. The surfaces are
defined by the functions $\bigl(U(R),V(R)\bigr)$:
\begin{equation}
  U = - R - 2M\ln\left(\frac{R}{2M} - 1\right) + T_\infty
\label{Uinfty}
\end{equation}
and 
\begin{equation}
  V = w + 4M\kappa\left[\exp\left(\frac{T_\infty +
  R}{4M}\right)\sqrt{\frac{R}{2M} - 1}\right]\ .
\label{Vinfty}
\end{equation}
The Schwarzschild time interval is the difference $T_{\infty 2} - T_{\infty
  1}$.  Again Eqs.~(\ref{84.1})--(\ref{Vinfty}) not only imply that the
intervals as calculated from $\tilde{V}$ or $T_\infty$ match with intervals
calculated from the geometry of $\bigl({\mathcal M}, \Psi_+(M,w;U,V)\bigr)$.
They also represent choices of origins, i.e., asymptotic frames, at ${\mathcal
  I}^-(+1,M,w)$ and ${\mathcal I}^0(+1,M,w)$, for each solution
$\Psi_+(M,w;U,V)$. 

The relation between the origins of the three spaces ${\mathcal I}^+$,
${\mathcal I}^-$ and ${\mathcal I}^0$ corresponding to {\em one} Schwarzschild
spacetime (i.e., for a fixed $M$) can be set by a geometric convention. For
example, the three surfaces $\tilde{U} = 0$, $\tilde{V} = 0$ and $T_\infty =
0$ (if the coordinates $\tilde{U}$, $\tilde{V}$ and $T_\infty$ are extended
into the spacetime as null or maximal surfaces) can be required to intersect
all at a sphere of radius $R = 2M\kappa(1)$. This is the convention we use for
each $\eta$, $M$ and $w$. In this way, the retarded arrival time $w$ can also
be calculated by a standardized way from the description of the dynamics of
the shell in an extended frame associated with ${\mathcal I}^-$ and ${\mathcal
  I}^0$. This is important for the action of time shifts as well as for the
calculation of the Liouville form. The dependence of the geometric convention
on the parameters $\eta$ and $M$ influences the form of asymptotic action of
dilatations and the time reversal.

The attachment of  ${\mathcal I}^+(-1,M,w)$, ${\mathcal I}^-(-1,M,w)$ and
${\mathcal I}^0(-1,M,w)$ to the spacetimes $\bigl({\mathcal M},
\Psi_-(M,w;U,V)\bigr)$ is entirely analogous and can be skipped.

We observe that the scries have not, in general, a fixed position in the
background manifold $\mathcal M$. This is due to the gauge fixing $\sigma$.
Any gauge fixing implies also a definition of points (events) of $\mathcal M$
by some of their geometrical properties. For example, at the point $(U,V)$,
two light-like surfaces meet that have a particular geometrical meaning. The
point at scri can also be defined by some geometrical properties. The latter
definition does not coincides with a limit of the former in our case. This
does not lead to any problems, if one does not try to use the definition of
points inside $\mathcal M$ in an improper way: it is not gauge invariant and
has not much physical meaning.

On the other hand, we can use the gauge fixing to calculate a lot of useful,
gauge invariant results. For example, we can determine how the group $G_c$
acts on the scri. Clearly, $G_c$ acts there only via $G_b$ so that, for
example, $G_0$ must act trivially, leaving all points of scri invariant. For
general $\varphi \in G$, a simple calculation reveals the following pattern.
Let the diffeomorphism $\varphi$ acts on the physical phase space as follows
\[
  \varphi : (\eta,M,w) \mapsto (\eta,M',w')\ ,
\]
that is, $\varphi$ does not contain any time reversal. Then
\begin{equation}
  \varphi_b : {\mathcal I}^+(\eta,M,w) \mapsto {\mathcal I}^+(\eta,M',w')
\label{60/1}
\end{equation}
and the map is a {\em bijection} between the two scries. The same claim holds
for ${\mathcal I}^-$ and ${\mathcal I}^0$. For example, the scri ${\mathcal
  I}^-(+1,M,w)$ has a past end point at $\partial_-{\mathcal M}$ with the
value $V_{\text{past}}$ of the coordinate $V$, $V_{\text{past}} = w + 4M$. The
time shift $\varphi_H(t)$ sends $V$ to $V + t$, hence $V_{\text{past}} \mapsto
V_{\text{past}} + t$, and this is the past end point of ${\mathcal I}^-(+1,M,w
+ t)$.

Similarly, the time reversal
\begin{equation}
  I : {\mathcal I}^+(\eta,M,w) \mapsto {\mathcal I}^-(-\eta,M,-w)\ ,
\label{60/2}
\end{equation}
etc., defines a bijection between the two scries.

Actually, we are going to identify all ${\mathcal I}^+$'s considering the
points with the same value of the coordinate $\tilde{U}$ as equal. This gives a
common ${\mathcal I}^+$ for all solutions. Similarly, common ${\mathcal I}^-$
and ${\mathcal I}^0$ can be defined. This structure is very important for the
interpretation of the theory. We assume that observers are living at this
common scri and are using the common frames (origins of time).

The definition of the common scri given above, together with the equations
such as (\ref{60/1}) and (\ref{60/2}), imply that the group $G_c$ acts on the
common scri. This action is not difficult to calculate. The result is:
\begin{eqnarray*}
  \varphi_H(t) & : & (\tilde{U},\tilde{V},T_\infty) \mapsto
  (\tilde{U}+t,\tilde{V}+t,T_\infty+t)\ , \\
  \varphi_D(s) & : & (\tilde{U},\tilde{V},T_\infty) \mapsto
  (\tilde{U}e^s,\tilde{V}e^s,T_\infty e^s)\ , \\
    \varphi_H(t) & : & (\tilde{U},\tilde{V},T_\infty) \mapsto
  (-\tilde{V},-\tilde{U},-T_\infty)\ . 
\end{eqnarray*}
The fact that the action of the time shift by $t$ shifts all coordinates
$\tilde{U}$, $\tilde{V}$ and $T_\infty$ by the same amount is the consequence
of the judicious choice of origins. Similarly, the homogeneous action of the
dilatations is due to correlations of our choices of frames for all different
values of the parameter $M$.

Clearly, the group $G_c$ acts transitively on $\Gamma$. If the
symplectic structure of the physical phase space $\Gamma$ were known, the
functions generating the infinitesimal transformations of the group could be
found. Then the dynamics based on the time shifts could be constructed.

\subsection{Canonical theory}
The variables $\eta$, $M$ and $w$ can play the role of Dirac
observables, but we do not know what are their Poisson brackets. In the
present section, we calculate the brackets. We start from a Hamiltonian
action principle for null shells and reduce it to physical phase space
applying the technique invented by Kucha\v{r}.

\subsubsection{The action}
As a Hamiltonian action principle that implies the dynamics of our system, we
take the action Eq.\ (2.6) of \cite{L-W-F} (see also \cite{K-W}). Let us
briefly summarize the relevant formulae. The spherically symmetric metric is
written in the form:
\begin{equation}
  ds^2 = -N^2d\tau^2 + \Lambda^2(d\rho + N^{\rho}d\tau)^2 + R^2d\Omega^2\ ,
\label{metric}
\end{equation}
the shell is described by its radial coordinate $\rho = {\mathbf
  r}$ and its conjugate momentum $\mathbf p$. The action reads
\begin{equation}
  S_0 = \int d\tau\left[{\mathbf p}\dot{\mathbf r} + \int
  d\rho(P_\Lambda\dot{\Lambda} + P_R\dot{R} - H_0)\right]\ ,
 \label{S0}
\end{equation}
and the Hamiltonian is
\[
  H_0 = N{\mathcal H} + N^\rho{\mathcal H}_\rho\ + N_\infty E_\infty\ ,
\] 
where $N$ and $N^\rho$ are the lapse and shift functions, $\mathcal H$ and
${\mathcal H}_\rho$ are the constraints,
\begin{eqnarray}
 {\mathcal H} & = & \text{G}\left(\frac{\Lambda P_\Lambda^2}{2R^2} -
 \frac{P_\Lambda P_R}{R}\right)  + \frac{1}{\text{G}}\left(
 \frac{RR''}{\Lambda} - \frac{RR'\Lambda'}{\Lambda^2} 
 + \frac{R^{\prime 2}}{2\Lambda} - \frac{\Lambda}{2}\right) \nn \\
 && +\frac{\eta{\mathbf p}}{\Lambda}\delta(\rho - {\mathbf r})\ ,
\label{LWF-H} \\
  {\mathcal H}_\rho & = & P_RR' - P_\Lambda'\Lambda - {\mathbf
 p}\delta(\rho - {\mathbf r})\ ,
\label{LWF-Hr}
\end{eqnarray}
and the prime or dot denote the derivatives with respect to $\rho$ or $\tau$.
The term $N_\infty E_\infty$ is the remainder of ${\mathcal
  H}_\infty[\vec{N}_\infty]$ (see Sec.~\ref{sec:phs}) in the case of
rotational symmetry. 
$N_\infty := \lim_{\rho \rightarrow \infty} N^{\rho}(\rho)$ and $E_\infty$
is the ADM mass (see \cite{L-W-F}). In the Schwarzschild spacetime with mass
parameter $M$ and asymptotic Schwarzschild time $T_\infty$, it holds that
\[
  N_\infty E_\infty = \frac{1}{\text{G}}M\dot{T}_\infty\ .
\]
The term can then be transfered from the Hamiltonian to the part of action
containing time derivatives---the so-called {\em Liouville form}, see
\cite{kuchS}. 

The Hamiltonian constraint function $\mathcal H$ is written in a way that
differs from \cite{L-W-F} and \cite{H-Kie} in that its dependence on the
Newton constant G becomes visible. This is necessary in order that the scaling
properties\footnote{To reveal the scaling behaviour, units must be chosen so
  that $\hbar = c = 1$, but G $\neq 1$!} of $\mathcal H$ are manifest,
${\mathcal H} \mapsto e^{-s}{\mathcal H}$. Observe that the rescaling of
$\mathcal H$ cannot be implemented by a canonical transformation: all function
that might be tried to generate such rescaling had vanishing Poisson brackets
with G. This does not mean, however, that vacuum Einstein equations are not
invariant under dilatation (cf.~\cite{A-T}). In our case, the geometric sector
alone (see Sec.~\ref{sec:mspace}) as well as the matter sector alone, are
invariant under dilatation, but the whole theory is not: the relation between
geometry and matter involves the Newton constant.

The ``volume'' variables $\Lambda$, $R$, $P_\Lambda$, $P_R$, $N$ and
${N}^\rho$ are the same as in \cite{L-W-F} and \cite{H-Kie}. The meaning of
the variables $\Lambda$, $R$, $N$ and ${N}^\rho$ can be inferred from the
spacetime metric (\ref{metric}). The momenta conjugate to the configuration
variables $\Lambda$ and $R$ can be calculated from the action $S_0$ by varying
it with respect to $P_\Lambda$ and $P_R$:
\begin{equation}
  P_\Lambda = -\frac{R}{\text{G}N}(\dot{R} - {N}^\rho R')\ ,
\label{L2.5b}
\end{equation}
and
\begin{equation}
  P_R = -\frac{\Lambda}{\text{G}N}(\dot{R} - {N}^\rho R') -
  \frac{R}{\text{G}N}[\dot{\Lambda} - ({N}^\rho\Lambda)']\ .
\label{L2.5c}
\end{equation}

Some differentiability conditions at the shell are important in order that
proper equations of motion are obtained by varying the action. One can assume
as in \cite{L-W-F} that the gravitational variables are smooth functions of
$\rho$, with the exception that ${N}'$, $({N}^\rho)'$, $\Lambda'$, $R'$,
$P_\Lambda$ and $P_R$ may have finite discontinuities at isolated values of
$\rho$. The coordinate loci of the discontinuities are smooth functions of
$\tau$ for each shell. This follows from (i) the conditions at shell points
which are the same as in \cite{L-W-F} and (ii) from the corresponding choice
of foliation: the metric with respect to coordinates $\tau$ and $\rho$ may be
piecewise smooth and everywhere continuous.

\subsubsection{The Liouville form at the constraint surface}
\label{sec:Liouville}
Our aim is to calculate the Poisson brackets between Dirac observables such as
$M$ and $w$. We can employ the property of the pull-back ${\Theta}_{\mathcal
  C}$ of the Liouville form ${\Theta}$ to the constraint surface $\mathcal C$
that it depends only on the Dirac observables as it has been explained in
Sec.~\ref{sec:gauge} (see also \cite{B-O}, \cite{H1} and \cite{H2}).  Its
external differential then defines the symplectic form of the physical phase
space, which determine the brackets. Observe further that the pull-back
${\Theta}_{\mathcal C}$, if integrated over $\tau$, gives the action of the
reduced system. Indeed, if one solves the constraints, only the Liouville form
remains from the action $S_0$.

Thus, we have to transform the Liouville form to the variables $M$, $w$ and a
set of observable-dependent embeddings; these variables form a coordinate
system on the constraint surface for each case $\eta = +1$ and $\eta =-1$.

An important point is to specify the family of embeddings that will be used.
The embeddings are given by
\[
  U = U(o,\rho),\quad V = V(o,\rho)\ ,
\]
where $U$ and $V$ are the coordinates defined in Sec.~\ref{sec:sol}. These
functions have to satisfy several conditions.
\begin{enumerate}
\item As $\Sigma$ is space-like, $U$ and $V$ are null and increasing towards
  the future, we must have $U' < 0$ and $V' > 0$ everywhere.
\item At the regular centre, the four-metric is flat and the three-metric is
  to be smooth. This implies, for all $o$'s, that $U'(o,0) = -V'(o,0)$ in
  addition to the condition $U(o,0) = V(o,0)$. This follows from
  $T'(U(o,0),V(o,0))=0$ and means that $\Sigma$ must run parallel to $T =$
  const in order to avoid conical singularities. Here, $T = (U + V)/2$ is the
  time of the inertial system at the centre.
\item At the space-like infinity, the four-metric is the Schwarzschild metric.
  We require that the embedding approaches the Schwarzschild-time-constant
  surfaces $T =$ const, and that $\rho$ becomes the Schwarzschild curvature
  coordinate $R$ asymptotically. More precisely, the behaviour of the
  Schwarzschild coordinates $T$ and $R$ along each embedding
  $U(o,\rho),V(o,\rho)$ must satisfy
  \begin{eqnarray}T(\rho) & = & T_\infty + O(\rho^{-1}),
  \label{Tinf} \\
  R(\rho) & = & \rho + O(\rho^{-1})\ .
  \label{Rinf}
  \end{eqnarray}
  The asymptotic coordinate $T_\infty$ is a gauge-invariant quantity and it
  possesses the status of an observable. It follows then from
  Eqs.~(\ref{Uinfty}) and (\ref{Vinfty}) that the embeddings must depend on
  $\eta$, $M$ and $w$. 
\item At the shell ($\rho = {\mathbf r}$) we require $U(o,\rho)$ and
  $V(o,\rho)$ to be $C^\infty$ functions of $\rho$. In fact, as the four-metric
  is continuous in the coordinates $U$ and $V$, but not smooth, only the
  $C^1$-part of this condition is gauge invariant. Jumps in all higher
  derivatives are gauge dependent, but the condition will simplify equations
  without influencing results.
\end{enumerate}

The Liouville form of the action (\ref{81.1}) can be written as follows:
\begin{equation}
  \Theta = {\mathbf p}\dot{\mathbf r} -
  \frac{1}{\text{G}}M \dot{T}_\infty + \int_0^\infty
  d\rho\,(P_\Lambda\dot{\Lambda} + P_R\dot{R})  \, . 
\label{4,1}
\end{equation}
We can now start to transform (\ref{4,1}) into Kucha\v{r} variables. It is
advantageous to let first the double-null coordinates $U$ and $V$ arbitrary
and the Dirac observables $o^i$, $i = 1,\dots,2n$ unspecified.  We just need
to know that the metric and the embeddings depends on $o^i$:
\[
  A = A(U,V;o)\ ,\quad R = R(U,V;o)\ ,\quad U = U(o,\rho)\ ,\quad V =
  V(o,\rho)\ .
\]
For any double-null gauge, the equations
\begin{eqnarray}
  4RR_{,UV} + 4R_{,U}R_{,V} + A & = & 0\ ,
\label{5,1}  \\
  AR_{,UU} - A_{,U}R_{,U} & = & 0\ ,
\label{5,2} \\
  AR_{,VV} - A_{,V}R_{,V} & = & 0
\label{5,3}
\end{eqnarray}
represent the condition that the transformation is performed at the constraint
surface $\Gamma$ (cf.~\cite{H-Kie}, Eqs.~(32)--(34)).

The transformation formulae for $\Lambda$, $R$, $N$ and $N^\rho$ can be read
off the metric:
\begin{eqnarray}
  R & = & R\bigl(U(o,\rho),V(o,\rho),o\bigr)\ ,
\label{0*} \\
  \Lambda & = & \sqrt{-AU'V'}\ , 
\label{1*} \\
  {\mathcal N} & = & -\frac{\dot{U}V'-\dot{V}U'}{2U'V'}\sqrt{-AU'V'}\ , 
\label{2*} \\
  {\mathcal N}^\rho & = & \frac{\dot{U}V'+\dot{V}U'}{2U'V'}\ .
\label{3*}
\end{eqnarray}
The transformation of the momenta $P_R$ and $P_\Lambda$ are obtained from
their definitions (\ref{L2.5b}) and (\ref{L2.5c}), into which
Eqs.~(\ref{0*})--(\ref{3*}) are substituted:
\begin{eqnarray}
  \text{G}P_\Lambda & = & \frac{R}{\sqrt{-AU'V'}}\left(R_{,U}U' -
  R_{,V}V'\right),  
\label{plambda} \\
  \text{G}P_R & = & R_{,U}U' - R_{,V}V' + \frac{RA_{,U}}{2A}U' -
  \frac{RA_{,V}}{2A}V' + 
  \frac{R}{2}\frac{U''}{U'} - \frac{R}{2}\frac{V''}{V'}\ .
\label{pr}
\end{eqnarray}  

The transformation of the shell variable $\mathbf r$ follows from the obvious
relations 
\[
  U(o,{\mathbf r}) = w\ ,\quad  V(o,{\mathbf r}) = w\ ,
\]
each valid in one of the cases $\eta = +1$ or $\eta = -1$. Finally, the value
of the momentum $\mathbf p$ at the constraint surface can be found if the
differentiability properties of the volume functions at the shell are used and
the coefficients at the $\delta$-functions in the constraints (\ref{LWF-H})
and (\ref{LWF-Hr}) are compared (for details see \cite{H-Kie}):
\[
  {\mathbf p} = -\eta R\Delta_{\mathbf r}(R') = -\text{G}\Lambda\Delta_{\mathbf
  r}(P_\Lambda)\ ,
\]
where $\Delta_{\mathbf r}(f)$ denotes a jump in the function $f$ across the
shell at the point $\rho = {\mathbf r}$.

The form (\ref{4,1}) can be divided into a boundary part (the first two terms
on the right-hand side) and the volume parts. Each volume part is associated
with a particular component of the space cut out by the shell; it has the form
\begin{equation}
  \int_a^b d\rho\,(P_\Lambda\dot{\Lambda} + P_R\dot{R})\ ,
\label{vol}
\end{equation}
where $a$ and $b$ are values of the coordinate $\rho$ at the boundary of the
volume. There are only two cases, $a = 0$, $b = {\mathbf r}$ and $a =
{\mathbf r}$, $b = \infty$.

Since the pull-back of the Liouville form cannot depend on the volume
variables $U(o,\rho)$ and $V(o,\rho)$, its volume part must be, after the
transformation, given by
\begin{equation}
  \Theta_a^b\vert_{\mathcal C} = \int_a^b d\rho\,[(f\dot{U} + g\dot{V} +
  h_i\dot{o}^i)' + \dot{\varphi}]\ ,
\label{ans}
\end{equation}
cf.~\cite{H-Kie}. Comparing the coefficients at the dotted and primed
quantities in Eqs.~(\ref{vol}) and (\ref{ans}) leads to partial differential
equations for the functions $f$, $g$ and $h_i$ \cite{H-Kie} and one can show
that Eqs.~(\ref{5,1})--(\ref{5,3}) are the integrability conditions for these
partial differential equations.

In such a way, $\Theta_{\mathcal C}$ can be transformed to a sum of boundary
terms. These terms can be calculated and simplified if the transformation
equations for the shell variables and the properties of the embeddings at the
boundary points are used. It is a lengthy and rather technical calculation
that is not very interesting for us now; it has been described in detail in
\cite{H-Kie} and, especially, \cite{H-Kou}. We skip it and quote only the
final results:
\begin{equation}
  \eta = +1:\quad \Theta\vert_{\mathcal C} = -\frac{1}{\text{G}}M\dot{w}\ ;
  \quad \eta = -1:\quad  \Theta\vert_{\mathcal C} = -\frac{1}{\text{G}}
  M\dot{w}\ . 
\label{Theta-end}
\end{equation}
Only one aspect of the calculation deserves mentioning. This is the cancellation
of the term $-(M/{\text{G}})\dot{T}_\infty$ in the Liouville form (\ref{4,1})
by a term coming from the volume part (\ref{vol}). The cancellation is again a
result of our judicious choice of asymptotic frames as described in
Sec.~\ref{sec:mspace}: the variable $T_\infty$ enters the calculation via
Eqs.~(\ref{Uinfty}) and (\ref{Vinfty}) that are implied by the choice.

At this stage, it is advantageous to introduce new variables for the
description of the Dirac observables, that is, for the matter sector in our
case: let us define, for $\eta = +1$:
\[
  p_u := -\frac{M}{\text{G}}\ ,\quad u := w\ ,
\]
and for $\eta = -1$,
\[
  p_v := -\frac{M}{\text{G}}\ ,\quad v := w\ .
\]
Observe that the momenta $p_u$ and $p_v$ have the meaning of negative energy
and scale in accord with this meaning as $e^{-s}$. The rescaling of $p$'s,
together with the rescaling by $e^s$ of $u$ and $v$, can be implemented
canonically.
 
The desired Poisson brackets can be written down immediately:
\[
  \{u,p_u\} = 1\ ,\quad \{v,p_v\} = 1\ .
\]
the generator of the infinitesimal time shift is $p_udt$ or $p_vdt$ and that of
dilatation is $p_uuds$ or $p_vvds$.

In \cite{H-Kou}, a general method of integration of the differential equations
for the functions $f$, $g$ and $h_i$ has been developed and a formula
generalizing Eq.~(\ref{Theta-end}) to a system containing any number of
in-going and any number out-going shell and the due shell intersections
(Kouletsis formula) has been written down.

The reduced action that contains only the physical degrees of freedom is, if
$\eta = +1$,
\begin{equation}
  S_+ = \int d\tau p_u\dot{u}\ ,
\label{actu}
\end{equation}
and if $\eta = -1$,
\begin{equation}
  S_- = \int d\tau p_v\dot{v}\ .
\label{actv}
\end{equation}
These actions imply the equations of motion
\[
  p_u = \text{const}\ ,\quad v = \text{const}
\]
or 
\[
  p_v = \text{const}\ ,\quad v = \text{const}\ ,
\]
as they are to be.

\subsubsection{Merging in- and out-going dynamics}
In this subsection, the two reduced action principles for the out-going and
in-going dynamics of the shell will be recognized as a result of reduction of
a single merger action. The merger action gives a complete account of all
motions of the shell on the background manifold $\mathcal M$. The merging is a
step of certain significance for our quantum theory.

Consider the action (\ref{actu}) for out-going shells. The variables $p_u$ and
$u$ do not contain the full information about where the shell is in $\mathcal
M$. The function $u(\tau)$ is the value of only one coordinate $U$ of the
shell for the value $\tau$ of the time parameter. For a complete description,
$v(\tau)$ had to be added. For a solution trajectory, $u(\tau)$ is constant,
but $v(\tau)$ can be arbitrary: it depends on the choice of the parameter
$\tau$ and so it can be considered as a gauge variable in a
reparametrization-invariant formalism. Hence, a valid extension of the action
(\ref{actu}) is
\[
  \bar{S}_+ = \int d\tau (p_u\dot{u} + p_v\dot{v} - n_+p_v)\ ,
\]
where $n_+$ is a Lagrange multiplier that enforces the vanishing of the
momentum $p_v$ conjugate to the gauge variable $v$. The meaning of $n_+$
transpires from the equation
\[
  \dot{v} = n_+
\]
obtained by varying $\bar{S}_+$ with respect to $p_v$. Thus, $n_+$ measures
the rate of $\tau$ with respect to the parameter $2t$: for any out-going
shell, we have
\[
  t = \frac{1}{2}\bigl(u + v(\tau)\bigr)\ ,
\]
where $t$ is the time of the inertial frame defined by the regular centre $R =
0$ and, at the same time, it coincides with the parameter of the time-shift
group.

We can define another Lagrangian multiplier $n$ by
\[
 n_+ = np_u\ ,
\]
because $p_u$ is always non-zero in the case $\eta = +1$. The multiplier $n$
measures the rate of $\tau$ with respect to the ``physical parameter'': the
corresponding tangent vector to particle trajectory coincides with the
contravariant four-momentum of the particle.

An analogous extension $\bar{S}_-$ of $S_-$, Eq.~(\ref{actv}) is
\[
  \bar{S}_- = \int d\tau (p_u\dot{u} + p_v\dot{v} - n_-p_u)\ ,
\]
and we can switch to $n$ defined by $n_- = np_v$ now.

The merger action is
\begin{equation}
  \bar{S} = \int d\tau (p_u\dot{u} + p_v\dot{v} - np_up_v)\ .
\label{merger}
\end{equation}
It contains both cases as the two possible solutions of the constraint
\[
  p_up_v = 0\ :
\]
if $p_v = 0$, we obtain the case $\eta = +1$, and if $p_u = 0$, $\eta = -1$. 

The two variables $u$ and $v$ determine the position of the shell in $\mathcal
M$ uniquely. The generator of the time translation is $p_u + p_v$, the value
of which is the negative of the total energy $E = M/$G of the system, and the
generator of the dilatation is $p_uu + p_vv$.

The new phase space has non-trivial boundaries:
\begin{equation}
  p_u \leq 0,\quad p_v \leq 0\ ,
\label{77}
\end{equation}
\begin{equation}
 \frac{-u+v}{2} > 0\ ,
\label{78}
\end{equation}
\begin{equation}
  p_v = 0\ ,\quad U \in (-\infty,u)\ , \quad V > u -
  4p_u\kappa\left(-\exp\frac{u-U}{4p_u}\right)\ , 
\label{79}
\end{equation}
and 
\begin{equation}
  p_u = 0\ ,\quad V \in (v,\infty)\ , \quad U < v +
  4p_v\kappa\left(-\exp\frac{V-v}{4p_v}\right)\ . 
\label{80}
\end{equation}
The boundaries defined by inequalities (\ref{79}) and (\ref{80}) are due
to the singularity. Eqs.~(\ref{79}) and (\ref{80}) limit the values of $U$ and
$V$ that can be used for embeddings.

The two dynamical systems defined by the actions (\ref{S0}) and
(\ref{merger}) are equivalent: each maximal dynamical trajectory of the
first, if transformed to the new variables, gives a maximal dynamical
trajectory of the second and vice versa.

The variables $u$, $v$, $p_u$ and $p_v$ span the effective phase space of the
shell. They contain all true degrees of freedom of the system. One can observe
that the action (\ref{merger}) coincides with the action for free motion of a
zero-rest-mass spherically symmetric (light-like) shell in flat spacetime.
Such a dynamics is complete because there is no geometric singularity at the
value zero of the radius of the shell, $(-u+v)/2$, and this point can be
considered as a harmless caustic so that the light can re-expand after passing
through it. The dynamics of the physical degrees of freedom by itself is,
therefore, regular. 

It might seem possible to extend the phase space of the gravitating shell,
too, in the same way so that the in-going and the out-going sectors are merged
together into one bouncing solution.  However, such a formal extension of the
dynamics (\ref{merger}) is not adequate. The physical meaning of any solution
written in terms of new variables is given by measurable quantities of
geometrical or physical nature, which include now also the curvature of
spacetime. These observables must be expressed as functions on the phase
space. They can of course be transformed between the phase spaces of the two
systems (\ref{S0}) and (\ref{merger}). They cannot be left out from any
complete description of a system, though they are often included only tacitly:
an action alone does not define a system. This holds just as well for the
action (\ref{S0}) as for (\ref{merger}).

Let us consider these observables. Eqs.\ (\ref{80.2})--(\ref{80.5}) and
(\ref{80.2-})--(\ref{80.5-}) can be used to show that the curvature of the
solution spacetime diverges at the boundary defined by Eq.\ (\ref{79}) for
$p_v = 0$ and by Eq.\ (\ref{80}) for $p_u = 0$. It follows that the observable
quantities at and near the ``caustic'' are badly singular and that there is no
sensible extension of the dynamics defined by action (\ref{merger}) to it, let
alone through it. This confirms the more or less obvious fact that no
measurable property (such as the singularity) can be changed by a
transformation of variables.

The action for the null dust shell is now written in a form which can be taken
as the starting point for quantization. Surprisingly, it will turn out that a
quantum theory can be constructed so that it is singularity-free.  This will
be shown in the next section.

\section{Quantum theory}
\label{sec:quantum}
In this section, we shall construct a quantum theory of our model. Of course,
there is no unique construction of a quantum theory for any given classical
model. We just show that there is a singularity-free, unitary quantum
mechanics of the shell.  The account in this section follows \cite{unitar}
adding a few new ideas.

\subsection{Group quantization}
\label{sec:quant}
To quantize the system defined by the action (\ref{merger}), we apply the
so-called group-theoretical quantization method \cite{isham}. There are three
reasons for this choice. First, the method as modified for the generally
covariant systems by Rovelli \cite{rovel} (see also \cite{honnef} and
\cite{H-I}) is based on the algebra of Dirac observables of the system;
dependent degrees of freedom don't influence the definition of Hilbert space.
Second, the group method has, in fact, been invented to cope with restrictions
such as Eqs.\ (\ref{77}) and (\ref{78}). By and large, one has to choose a set
of observables that form a Lie algebra; this algebra has to generate a group
of symplectomorphisms that has to act transitively in the phase space
respecting all boundaries. In this way, the information about the boundaries
in built in the quantum mechanics. Finally, the method automatically leads to
self-adjoint operators representing all observables. In particular, a
self-adjoint extension of the Hamiltonian is obtained in this way, and this is
the reason that the dynamics is unitary.

To begin with, we have to find a complete set of Dirac observables. Let us
choose the functions $p_u$, $p_v$, $D_u :=up_u$ and $D_v :=vp_v$. Observe that
$u$ alone is constant only along out-going shell trajectories ($p_u \neq 0$),
and $v$ only along in-going ones ($p_v \neq 0$), but $up_u$ and $vp_v$ are
always constant. The only non vanishing Poisson brackets are
\[
  \{D_u,p_u\} = p_u,\quad \{D_v,p_v\} = p_v.
\]
This Lie algebra generates a group $Q_2$ of symplectic transformations of the
phase space that preserve the boundaries $p_u = 0$ and $p_v=0$. $Q_2$ is the
Cartesian product of two copies of the affine group $\mathcal A$ on $\mathbb
R$.

The group $\mathcal A$ generated by $p_u$ and $D_u$ has three
irreducible unitary representations. In the first one, the spectrum of
the operator $\hat{p}_u$ is $[0,\infty)$, in the second, $\hat{p}_u$
is the zero operator, and in the third, the spectrum is $(-\infty,0]$, see
Ref.\ \cite{raczka}. Thus, we must choose the third representation; this can
be described as follows (details are given in Ref.\ \cite{raczka}).

The Hilbert space is constructed from complex functions $\psi_u(p)$
of $p\in [0,\infty)$; the scalar product is defined by
\[ 
  (\psi_u,\phi_u) := \int_0^\infty\ \frac{dp}{p}\ \psi^*_u(p)\phi_u(p),
\]
and the action of the generators $\hat{p}_u$ and $\hat{D}_u$ on smooth
functions is
\[
  (\hat{p}_u\psi_u)(p) = -p\psi_u(p),\quad (\hat{D}_u\psi_u)(p) =
  -ip\frac{d\psi_u(p)}{dp}.
\]
Similarly, the group generated by $p_v$ and $D_v$ is represented on
functions $\psi_v(p)$; the group $Q_2$ can, therefore, be represented
on pairs $\Bigl(\psi_u(p),\psi_v(p)\Bigr)$ of functions:
\begin{eqnarray*}
\hat{p}_u\Bigl(\psi_u(p),\psi_v(p)\Bigr) & = & \Bigl(-p\psi_u(p),0\Bigr), \\ 
\hat{p}_v\Bigl(\psi_u(p),\psi_v(p)\Bigr) & = & \Bigl(0,-p\psi_v(p)\Bigr), \\
\hat{D}_u\Bigl(\psi_u(p),\psi_v(p)\Bigr) & = &
\Bigl(-ip\frac{d\psi_u(p)}{dp},0\Bigr), \\ 
\hat{D}_v\Bigl(\psi_u(p),\psi_v(p)\Bigr) & = &
\Bigl(0,-ip\frac{d\psi_v(p)}{dp}\Bigr)\ ,  
\end{eqnarray*}
This choice guarantees that the Casimir operator $\hat{p}_u\hat{p}_v$
is the zero operator on this Hilbert space, and so the constraint is
satisfied. The scalar product is
\[
  \biggl(\bigl(\psi_u(p),\psi_v(p)\bigr),
  \bigl(\phi_u(p),\phi_v(p)\bigr)\biggr) = \int_0^\infty \frac{dp}{p}
  \bigl(\psi^*_u(p)\phi_u(p) + \psi^*_v(p)\phi_v(p)\bigr)\ .
\]
Let us call this representation ${\mathcal R}_2$. Observe that this
representation of the group ${\mathcal A} \times {\mathcal A}$ is not
irreducible. It is, however, irreducible for the group extended by the time
reversal because its representative swaps the invariant subspaces. The
representation ${\mathcal R}_2$ has a well-defined meaning as a quantum theory
of our system.  In it, the out-going shells are independent of the in-going
ones. The dynamics of each kind of shells is complete for itself. One could
study this dynamics by defining a position operator $\hat{r}(t)$ in a natural
way similar to what will be done later and one would find that the in-going
shells simply proceed through $r = 0$ into negative values of $r$. Analogous
holds, in the time reversed order, for the out-going shells.

Handling the inequality (\ref{78}) is facilitated by the canonical
transformation:
\begin{alignat}{2}
  t & = (u+v)/2, & \qquad r & = (-u+v)/2,
\label{tr} \\
  p_t & = p_u + p_v, & \qquad p_r & = -p_u + p_v.
\label{ptpr}
\end{alignat}
The constraint function then becomes $p_up_v = (p_t^2 - p_r^2)/4$.

The function $p_t\delta t$ generates, via Poisson brackets, the infinitesimal
time shift in $(t,r)$-space, $t \mapsto t + \delta t$, $r \mapsto r$, and
$p_r\delta r$ generates an $r$-shift, $t \mapsto t$, $r \mapsto r + \delta
r$. We introduce also the observables
\[
  D := D_u + D_v = tp_t + rp_r\ ,
\]
and 
\[
  J := -D_u + D_v = rp_t + tp_r\ .
\]
The function $D\delta s$ generates a dilatation in the $(t,r)$-space, $t
\mapsto t + t\delta s$, $r \mapsto r + r\delta s$, and $J\delta v$
generates a boost, $t \mapsto t + r\delta v$, $r \mapsto r + t\delta
v$. What do these transformations with our half-plane $r > 0$? The
transformations generated by $p_t$ and $D$ preserve the boundary while those
by $p_r$ and $J$ do not. Hence, only the subgroup $Q_1$ of $Q_2$ generated by
$p_t$ and $D$ respects the inequality (\ref{78}).

The representation ${\mathcal R}_2$ of $Q_2$ can be decomposed into the direct
sum of two equivalent representations of $Q_1$. This representation of $Q_1$
will be denoted by ${\mathcal R}_1$ and it will serve as our {\em definitive}
quantum mechanics. Let us stress that this quantum mechanics does not describe
a subsystem of the physical system described with the representation
${\mathcal R}_2$. The representation ${\mathcal R}_1$ can be described as
follows: The states are determined by complex functions $\varphi(p)$ on
${\mathbb R}_+$; the scalar product $(\varphi,\psi)$ is
\[
  (\varphi,\psi) = \int_0^\infty\frac{dp}{p}\ \varphi^*(p)\psi(p);
\]
let us denote the corresponding Hilbert space by $\mathcal K$. The
representatives of the above algebra are
\begin{eqnarray*}
 (\hat{p}_t\varphi)(p) & = & -p\varphi(p), \\
 (\hat{p}_r^2\varphi)(p) & = & p^2\varphi(p), \\
 (\hat{D}\varphi)(p) & = & -ip\frac{d\varphi(p)}{dp}, \\
 (\hat{J}^2\varphi)(p) & = & -p\frac{d\varphi(p)}{dp} -
 p^2\frac{d^2 \varphi(p)}{dp^2}. 
\end{eqnarray*}
There are also well-defined operators for $p_r^2$ and $J^2$ because, in the
representation ${\mathcal R}_2$, the identities $\hat{p}_r^2 =
\hat{p}_t^2$ and $\hat{J}^2 = \hat{D}^2$ hold.

An important observation is that ${\mathcal R}_1$ is a representation of the
group of asymptotic symmetries as described in Sec.~\ref{sec:mspace}. Another
important observation is that the quantum mechanics ${\mathcal R}_2$ describes
two discoupled degrees of freedom, namely the in- and out-going shells, while
${\mathcal R}_1$ describes a system with a single degree of freedom: the in-
and out-going motions have been coupled into one motion. Let us study, what is
this motion.

First, we have to construct a time evolution. For that, we use the time shift
symmetry generated by $\hat{p}_t$ The operator $-\hat{p}_t$ has the meaning of
the total energy $E = M/$G of the system.  We observe that it is a
self-adjoint operator with a positive spectrum and that it is diagonal in our
representation. The parameter $t$ of the unitary group $\hat{U}(t)$ that is
generated by $-\hat{p}_t$ is easy to interpret: $t$ represents the quantity
that is conjugated to $p_t$ in the classical theory and this is given by Eq.\ 
(\ref{tr}). Hence, $\hat{U}(t)$ describes the evolution of the shell states,
for example, between the levels of the function $(U+V)/2$ on $\mathcal M$.

The missing piece of information of where the shell is on $\mathcal M$ is
carried by the quantity $r$ of Eq.\ (\ref{tr}). We shall define the
corresponding position operator in three steps.

First, we observe that $r$ itself is not a Dirac observable, but the boost $J$
is, and that the value of $J$ at the surface $t=0$ coincides with $rp_t$. It
follows that the meaning of the Dirac observable $Jp_t^{-1}$ is the position
at the time $t=0$. This is in a nice correspondence with the Newton-Wigner
construction \cite{N-W} on one hand, and with the notion of evolving constants
of motion by Rovelli \cite{Revolv} on the other.

Second, we try to make $Jp_t^{-1}$ into a symmetric operator on our Hilbert
space. However, in the representation ${\mathcal R}_1$, only $\hat{J}^2$ is
meaningful. Let us then chose the following factor ordering for $J^2p_t^{-2}$:
\begin{equation}
 \hat{r}^2 := \frac{1}{\sqrt{p}}\hat{J}\frac{1}{p}\hat{J}\frac{1}{\sqrt{p}} =
 -\sqrt{p}\frac{d^2}{dp^2}\frac{1}{\sqrt{p}}.
\label{rsym}
\end{equation}
Other choices are possible; the above one makes $\hat{r}^2$
essentially a Laplacian and this simplifies the subsequent
mathematics. Indeed, we can map $\mathcal K$ unitarily to
$L^2({\mathbb R}_+)$ by sending each function $\psi(p) \in {\mathcal
K}$ to $\tilde{\psi}(p) \in L^2({\mathbb R}_+)$ as follows:
\[
  \tilde{\psi}(p) = \frac{1}{\sqrt{p}}\psi(p).
\]
Then, the operator of squared position $\tilde{r}^2$ on $L^2({\mathbf
R}_+)$ corresponding to $\hat{r}^2$ is
\[
  \tilde{r}^2\tilde{\psi}(p) =
  \frac{1}{\sqrt{p}}\hat{r}^2\Bigl(\sqrt{p}\tilde{\psi}(p)\Bigr) 
  = - \frac{d^2\tilde{\psi(p)}}{dp^2} = -\tilde{\Delta}\tilde{\psi}(p).
\]

Third, we have to extend the operator $\hat{r}^2$ to a self-adjoint
one. The Laplacian on the half-axis possesses a one-dimensional family
of such extensions \cite{reed-s}. The parameter is $\alpha \in
[0,\pi)$ and the domain of $\tilde{\Delta}_\alpha$ is defined by the
boundary condition at zero:
\[
  \tilde{\psi}(0) \sin\alpha + \tilde{\psi}'(0) \cos\alpha = 0.
\]
The complete system of normalized eigenfunctions of $\tilde{\Delta}_\alpha$
can then easily be calculated:
\[
  \tilde{\psi}_\alpha(r,p) = \sqrt{\frac{2}{\pi}}\ \frac{r\cos\alpha\cos
  rp - \sin\alpha\sin rp}{\sqrt{r^2\cos^2\alpha + \sin^2\alpha}};
\]
if $\alpha \in (0,\pi/2)$, there is one additional bound state,
\[
  \tilde{\psi}_\alpha(b,p) = \frac{1}{\sqrt{2\tan\alpha}}\exp(-p\tan\alpha),
\]
so that
\begin{eqnarray*}
  -\tilde{\Delta}_\alpha\tilde{\psi}_\alpha(r,p) & = &
  r^2\tilde{\psi}_\alpha(r,p), \\
  -\tilde{\Delta}_\alpha\tilde{\psi}_\alpha(b,p) & = &
  -\tan^2\alpha\,\tilde{\psi}_\alpha(r,p).
\end{eqnarray*}
The corresponding eigenfunctions $\psi_\alpha$ of the operator
$\hat{r}^2_\alpha$ are:
\[
  \psi_\alpha(r,p) = \sqrt{\frac{2p}{\pi}}\ \frac{r\cos\alpha\cos
  rp - \sin\alpha\sin rp}{\sqrt{r^2\cos^2\alpha + \sin^2\alpha}},
\]
and we restrict ourselves to $\alpha \in [\pi/2,\pi]$, so that there are no
bound states and the operator $\hat{r}$ is self-adjoint. Indeed, $\tilde{r}$
is the square root of $-\tilde{\Delta}_\alpha$, hence its eigenvalue for the
bound state is imaginary.

To restrict the choice further, we apply the idea of Newton and Wigner
\cite{N-W}.  First, the subgroup of $Q_0$ that preserves the surface $t = 0$
is to be found. This is, in our case, $U_D(s)$ generated by the dilatation
$D$.  Then, in the quantum theory, the eigenfunctions of the position at $t=0$
are to transform properly under this group; this means that the eigenfunction
for the eigenvalue $r$ is to be transformed to that for the eigenvalue
$e^{-s}r$, for each $s$ and $r$.  The dilatation group generated by $\hat{D}$
acts on a wave function $\psi(p)$ as follows:
\[
  \psi(p) \mapsto U_D(s)\psi(p) = \psi(e^{-s}p),
\]
where $U_D(s)$ is an element of the group parameterized by
$s$. Applying $U_D(s)$ to $\psi_\alpha(r,p)$ yields
\[
  U_D(s)\psi_\alpha(r,p) =
  e^{-s/2}\sqrt{\frac{2p}{\pi}}\ \frac{r\cos\alpha\cos
  (e^{-s}rp) - \sin\alpha\sin
  (e^{-s}rp)}{\sqrt{r^2\cos^2\alpha + \sin^2\alpha}}.
\]
The factor $e^{-s/2}$ in the resulting functions of $p$
keeps the system $\delta$-normalized.

Let $\alpha = \pi/2$; then
\[
  U_D(s)\psi_{\pi/2}(r,p) = e^{-s/2}\psi_{\pi/2}(e^{-s}r,p).
\]
Similarly, for $\alpha = \pi$,
\[
  U_D(s)\psi_{\pi}(r,p) = e^{-s/2}\psi_{\pi}(e^{-s}r,p),
\]
but such relation can hold for no other $\alpha$ from the interval
$[\pi/2,\pi]$, because of the form of the eigenfunction dependence on $r$. Now,
Newton and Wigner require that 
\[
  U_D(s)\psi(r,p) = e^{-s/2}\psi(e^{-s}r,p).
\]  
Then all values of $\alpha$ except for $\alpha = \pi/2$ and $\alpha =
\pi$ are excluded.

We have, therefore, only two choices for the self-adjoint extension of
$\hat{r}^2$:
\begin{equation}
  \psi(r,p) := \sqrt{\frac{2p}{\pi}}\sin rp,\quad r \geq 0,
\label{rsa}
\end{equation}
and 
\[
  \psi(r,p) := \sqrt{\frac{2p}{\pi}}\cos rp,\quad r \geq 0.
\]
Let us select the first set, Eq.\ (\ref{rsa}); by that, the construction of a
position operator is finished.  The construction contains a lot of choice: the
large factor-ordering freedom, and the freedom of choosing the self-adjoint
extension.

Another observable that we shall need is $\hat{\eta}$; this is to tell us the
direction of motion of the shell at the time zero, having the eigenvalues $+1$
for all purely out-going shell states, and $-1$ for the in-going ones. In fact,
in the classical theory, $\eta = -\mathrm{sgn}p_r$, but $p_r$ does not act as
an operator on the Hilbert space $\mathcal K$, only $p_r^2$, and the sign is
lost. 

Consider the classical dilatation generator $D = tp_t + rp_r$. It is a Dirac
observable; at $t = 0$, its value is $rp_r$. Thus, for positive $r$, the sign
of $-D$ at $t = 0$ has the required value.  Hence, we have the relation:
\[
  \text{sgn}D = -\eta_{t=0}.
\]
 
The eigenfunctions $\psi_a(p)$ of the operator $\hat{D}$ are solutions of the
differential equation:
\[
  \hat{D}\psi_a(p) = a\psi_a(p).
\]
The corresponding normalized system is given by
\[
  \psi_a(p) = \frac{1}{\sqrt{2\pi}}p^{ia}.
\]
Hence, the kernels $P_\pm(p,p')$ of the projectors $\hat{P}_\pm$ on
the purely out- or in-going states are:
\[
  P_+(p,p') =
  \int_{-\infty}^0da\,\psi_a(p)\frac{\psi_a^*(p')}{p'},\quad P_-(p,p')
  = \int_0^\infty da\, \psi_a(p)\frac{\psi_a^*(p')}{p'}
\]
so that
\[
  (\hat{\eta}_0\psi)(p) = \int_0^\infty dp' [P_+(p,p') - P_-(p,p')]\psi(p').
\]

The observables $\hat{p}_t$, $\hat{D}$, $\hat{r}_0$ and $\hat{\eta}_0$ will
suffice to work out a number of interesting predictions.

\subsection{Motion of wave packets}
\label{sec:packets}
We shall work with the family of wave packets on the energy half-axis that are
defined by
\[
  \psi_{\kappa\lambda}(p) := \frac{(2\lambda)^{\kappa+1/2}}{\sqrt{(2\kappa)!}}
  p^{\kappa+1/2}e^{-\lambda p},
\]
where $\kappa$ is a positive integer and $\lambda$ is a positive number with
dimension of length. Using the formula
\begin{equation}
  \int_0^\infty dp\,p^ne^{-\nu p} = \frac{n!}{\nu^{n+1}},
\label{22'1}
\end{equation}
which is valid for all non-negative integers $n$ and for all complex
$\nu$ that have a positive real part, we easily show that the wave
packets are normalized,
\[
  \int_0^\infty\frac{dp}{p}\ \psi_{\kappa\lambda}^2(p) = 1.
\]

Observe that $\psi_{\kappa\lambda}(p) = \psi_{\kappa 1}(\lambda p)$. The
theory has been constructed so that it is scale invariant. The rescaling $p
\mapsto \lambda p$ does not change the scalar product.  The meaning of the
parameter $\lambda$ is, therefore, a scale: all energies concerning the packet
will be proportional to $\lambda^{-1}$, all times and all radii to $\lambda$.

The expected energy, 
\[
  \langle E\rangle_{\kappa\lambda} := \int_0^\infty\frac{dp}{p}\
  p\psi_{\kappa\lambda}^2(p),
\]
of the packet can be calculated by the formula (\ref{22'1}) with the simple
result
\[
 \langle E\rangle_{\kappa\lambda} = \frac{\kappa+1/2}{\lambda}. 
\]
The (energy) width of the packet can be represented by the mean quadratic
deviation (or dispersion), 
\[
  \langle\Delta E\rangle_{\kappa\lambda} := \sqrt{\langle
  E^2\rangle_{\kappa\lambda} - \langle E\rangle^2_{\kappa\lambda}}\ , 
\]
which is
\[
  \langle\Delta E\rangle_{\kappa\lambda} = \frac{\sqrt{2\kappa+1}}{2\lambda}.
\]

In the Schroedinger picture, the time evolution of the packet is generated by
$-\hat{p}_t$:
\[
  \psi_{\kappa\lambda}(t,p) = \psi_{\kappa\lambda}(p) e^{-ipt}.
\]
Let us calculate the corresponding wave function
$\Psi_{\kappa\lambda}(r,t)$ in the $r$-representation,
\[
 \Psi_{\kappa\lambda}(t,r) := \int_0^\infty\frac{dp}{p}\ 
 \psi_{\kappa\lambda}(t,p)\psi(r,p),
\]
where the functions $\psi(r,p)$ are defined by Eq.\
(\ref{rsa}). Formula (\ref{22'1}) then yields:
\begin{equation}
  \Psi_{\kappa\lambda}(t,r) = \frac{1}{\sqrt{2\pi}}
  \frac{\kappa!(2\lambda)^{\kappa+1/2}}{\sqrt{(2\kappa)!}}
  \left[\frac{i}{(\lambda +it +ir)^{\kappa+1}} - \frac{i}{(\lambda +it
  -ir)^{\kappa+1}}\right]. 
\label{17:1}
\end{equation}
It follows immediately that
\[
  \lim_{r\rightarrow 0}|\Psi_{\kappa\lambda}(t,r)|^2 = 0.
\]
The scalar product measure for the $r$-re\-pre\-sentation is just $dr$ because
the eigenfunctions (\ref{rsa}) are normalized, so the probability to find the
shell between $r$ and $r+dr$ is $|\Psi_{\kappa\lambda}(t,r)|^2dr$.

Our first important result is, therefore, that the wave packets start
away from the center $r=0$ and then are keeping away from it during
the whole evolution. This can be interpreted as the {\em absence of
singularity} in the quantum theory: no part of the packet is squeezed
up to a point, unlike the shell in the classical theory.

Observe that the equation $\Psi_{\kappa\lambda}(t,0) = 0$ is {\em not} a
result of an additional boundary condition imposed on the wave function. It
follows from the dynamics we have constructed. The nature of the question that
we are studying requires that the wave packets start in the asymptotic region
so that their wave function vanishes at $r = 0$ for $t \rightarrow -\infty$;
this is the only condition put in by hand. The fact that the dynamics
preserves this equation is the property of the self-adjoint extensions of the
Hamiltonian and the position operators.

Again, $\lambda$ can be used to re-scale the radius, $r = \lambda\rho$ and the
time, $t = \lambda \tau$ with the result
\[
  \Psi_{\kappa\lambda}(\lambda\tau,\lambda\rho) =
  \frac{1}{\sqrt{\lambda}}\Psi_{\kappa 1}(\tau,\rho)\ .
\]
The factor $1/\sqrt{\lambda}$ is due to the scalar product in the
$r$-representation being
\[
  \bigl(\Phi(r),\Psi(r)\bigr) = \int_0^\infty dr \Phi^*(r)\Psi(r) =
  \int_0^\infty d\rho \sqrt{\lambda}\Phi^*(\rho)\sqrt{\lambda}\Psi(\rho)\ ,
\]
so that $\sqrt{\lambda}\Psi(\lambda\rho)$ is $\lambda$-independent.

A more tedious calculation is needed to obtain the time dependence
$\langle r_t\rangle_{\kappa\lambda}$ of the expected radius of the shell,
\begin{equation}
  \langle r_t\rangle_{\kappa\lambda} := \int_0^\infty
  dr\,r|\Psi_{\kappa\lambda}(t,r)|^2.
\label{17:2}
\end{equation}

The results that can be calculated analytically are (we skip the calculations;
for details, see \cite{unitar}):
\[
  \langle r_t\rangle_{\kappa\lambda} = \langle r_{-t}\rangle_{\kappa\lambda}
  \quad \forall \kappa,\ \lambda\ ,t\ ,
\]
that is all packets are time reversal symmetric. There is the minimal
expected radius $\langle r_0\rangle_{\kappa\lambda}$ at $t=0$,
\begin{equation}
  \langle r_0\rangle_{\kappa\lambda} =
  \frac{1}{\pi}\frac{2^{2\kappa}(\kappa!)^2}{(2\kappa)!}
  \frac{\kappa+1}{\kappa}\frac{\lambda}{\kappa+1/2} > 0\ .
\label{r0}
\end{equation}

Let us turn to the asymptotics $t\rightarrow\pm\infty$. We obtain for both
cases $t\rightarrow \pm\infty$ and all $\lambda$ and $\kappa$:
\begin{equation}
  \langle r_t\rangle_{\kappa\lambda} \approx |t| + O(t^{-2\kappa}).
\label{rinfty}
\end{equation}

We can also calculate the spread of the wave packet in $r$ by means of the
$\hat{r}$-dispersion $\langle\Delta r_t\rangle_{\kappa\lambda}$. The
calculation of $\langle{r}^2\rangle_{\kappa\lambda}$ is much easier than that
of $\langle\hat{r}\rangle_{\kappa\lambda}$ because $\hat{r}^2$ is a
differential operator in $p$-representation:
\[
  \langle{r}^2\rangle_{\kappa\lambda} = -\lambda^2 \int_0^\infty
  \frac{dq}{q} \psi^*_{\kappa 1}(q,\tau)
  \left(-\sqrt{q}\frac{\partial^2}{\partial q^2} \frac{1}{\sqrt{q}}
  \psi_{\kappa 1}(q,\tau)\right)\ ,
\]
where $q = \lambda p$. This gives
\begin{equation}
  \langle{r}_t^2\rangle_{\kappa\lambda} = t^2 + \frac{\lambda^2}{2\kappa +
  1}\ .
\label{rsquar}
\end{equation}

For $\kappa \gg 1$, we can use the asymptotic expansion for the
$\Gamma$-function to obtain
\[
  \frac{2^{2\kappa+1}(\kappa!)^2}{(2\kappa)!} = \sqrt{2\pi(2\kappa+1)} -
  \frac{1}{4}\sqrt{\frac{2\pi}{2\kappa+1}} + O(\kappa^{-3/2})\ ;
\] 
that gives
\begin{equation}
  \langle r_0\rangle_{\kappa\lambda} =
  \lambda\left[\frac{1}{\sqrt{\pi\kappa}} + O(\kappa^{-3/2})\right]\ .
\label{rmin}
\end{equation}
Then, the spread of the packet for large $\kappa$ is
\begin{equation}
  \langle\Delta r_0\rangle_{\kappa\lambda} =
  \lambda\left[\sqrt{\frac{\pi-2}{\kappa}} + O(\kappa^{-3/2})\right]\ .
\label{deltarmin}
\end{equation}
We observe that the spread is of the same order as the expected value. At $t
\rightarrow \pm \infty$,
\begin{equation}
  \langle\Delta r_{\pm\infty}\rangle_{\kappa\lambda} =
  \lambda\left[\frac{1}{\sqrt{2\kappa+1}} + O(t^{-2\kappa+1})\right]\ .
\label{deltarinf}
\end{equation}
Hence, the asymptotic spread is nearly equal to the spread at the minimum.
This can be due to the light-like nature of the shell matter.

A further interesting question about the motion of the packets is about the
portion of a given packet that moves in---is purely in-going---at a given time
$t$. The portion is given by $\|\hat{P}_-\psi_{\kappa\lambda}\|^2$, where
$\hat{P}_-$ is the projector defined in Sec.\ \ref{sec:quant}. If we write out
the projector kernel and make some simple rearrangements in the expression of
the norm, we obtain:
\[
  \|\hat{P}_-\psi_{\kappa\lambda}\|^2 = \int_{-\infty}^\infty dq'
  \int_{-\infty}^\infty dq'' \left(\int_0^\infty
  da\,\psi^*_a(e^{q'})\psi_a(e^{q''})\right)
  \psi^*_{\kappa\lambda}(t,e^{q''}) \psi_{\kappa\lambda}(t,e^{q'}),
\]
where the transformation of integration variables $p'$ and $p''$ to $e^{q'}$
and $e^{q''}$ in the projector kernels has been performed. The further
calculation can be found in \cite{unitar}. The results are:
\[
  \|\hat{P}_-\psi_{\kappa\lambda}\|^2_{t=0} = 1/2\ ,\quad
  \|\hat{P}_-\psi_{\kappa\lambda}\|^2_{t\rightarrow -\infty} = 1\ ,\quad
  \|\hat{P}_-\psi_{\kappa\lambda}\|^2_{t\rightarrow \infty} = 0\ . 
\]
Hence, we have one-to-one relation between in- and out-going states at the
time of the bounce, while there are only in-going, or only out-going states at
the infinity.

The obvious interpretation of these formulae is that quantum shell always
bounces at the center and re-expands. 

The result that the quantum shell bounces and re-expands is clearly at odds
with the classical idea of black hole forming in the collapse and preventing
anything that falls into it from re-emerging. It is, therefore, natural to ask,
if the packet is squeezed enough so that an important part of it comes under
its Schwarzschild radius. We can try to answer this question by comparing the
minimal expected radius $\langle r_0\rangle_{\kappa\lambda}$ and its
spread  $\langle\Delta r_0\rangle_{\kappa\lambda}$ with the expected
Schwarzschild radius $\langle r_H\rangle_{\kappa\lambda}$  and its spread
$\langle\Delta r_H\rangle_{\kappa\lambda}$ for the wave packet. For the
Schwarzschild radius, we have
\[
  \langle r_H\rangle_{\kappa\lambda} = 2\text{G}\bar{M}_{\kappa\lambda} =
  (2\kappa+1)\frac{{L}^2_{\text{P}}}{\lambda}\ ,
\]
where ${L}_{\text{P}}$ is the Planck length, and
\[
  \langle\Delta r_H\rangle_{\kappa\lambda} = 2\text{G}\langle\Delta
  M\rangle_{\kappa\lambda} = \sqrt{2\kappa+1}\frac{{L}^2_{\text{P}}}{\lambda}\ .
\]
We are to ask the question: for which $\lambda$ and $\kappa$, the following
inequality holds:
\[
  \langle r_0\rangle_{\kappa\lambda} + \langle\Delta
  r_0\rangle_{\kappa\lambda} < \langle r_H\rangle_{\kappa\lambda} -
  \langle\Delta r_H\rangle_{\kappa\lambda}\ .
\]
If it holds, then the most of the packet becomes squeezed beyond the most of
the Schwarzschild radius. From Eqs.~(\ref{rmin}) and (\ref{deltarmin}), we
obtain
\[
  \frac{\lambda^2}{{L}^2_{\text{P}}}\left(\frac{1}{\sqrt{\pi}} + \sqrt{\pi -
  2}\right) < (2\kappa + 1)\sqrt{\kappa} - \sqrt{\kappa(2\kappa + 1)}\ .
\]
Clearly, if $\lambda \approx {L}_{\text{P}}$, then the inequality holds for any
large $\kappa$ (starting from $\kappa = 2$). For larger $\lambda$, $\kappa$
must be $(\lambda/{L}_{\text{P}})^{4/3} \times$ larger. Hence, we must go to
Planck regime in order that the packets cross their Schwarzschild radius.

To summarize: The packet can, in principle, fall under its Schwarzschild
radius. Even in such a case, the packet bounces and re-expands.

This apparent paradox will be explained in the next section.

\subsection{Grey horizons}
In this section, we try to explain the apparently contradictory result that
the quantum shell can cross its Schwarzschild radius in both directions. The
first possible idea that comes to mind is simply to disregard everything that
our model says about Planck regime. This may be justified, because the model
can hardly be considered as adequate for this regime. Outside Planck regime,
however, the shell bounces before it reaches its Schwarzschild radius and
there is no paradox. However, the model {\em is} mathematically consistent,
simple and solvable; it must, therefore, provide some mechanism to make the
horizon leaky. We shall study this mechanism in the hope that it can work in
more realistic situations, too.

To begin with, we have to recall that the Schwarzschild radius is the radius
of a non-diverging null hyper-surface; anything moving to the future can cross
such a hyper-surface only in one direction. The local geometry is that of an
apparent horizon. (Whether or not an {\em event} horizon forms is another
question; the answer to it also depends on the geometry near the singularity
\cite{H-contra}). However, as Einstein's equations are invariant under time
reversal, there are two types of Schwarzschild radius: that associated with a
black hole and that associated with a white hole. Let us call these
Schwarzschild radii (or apparent horizons) themselves {\em black} and {\em
  white}. The explanation of the paradox that follows from our model is that
quantum states can contain a linear combination of black and white (apparent)
horizons, and that no event horizon ever forms. We call such a combination a
{\em grey horizon}.

The existence of grey horizons can be shown as follows. The position and the
``colour'' of a Schwarzschild radius outside the shell is determined by the
spacetime metric. For our model, this metric is a combination of purely gauge
and purely dependent degrees of freedom, and so it is determined, within the
classical version of the theory, by the physical degrees of freedom through
the constraints.

The metric can be calculated along a Cauchy surface $\Sigma$ that intersect
the shell with total energy $E$, direction of radial motion $\eta$, and the
radius $r$. The result of this calculation (see \cite{unitar}) is: If the
shell is contracting, $\eta = -1$, then any space-like surface containing such
a shell can at most intersect an {\em out-going} apparent horizon at the
radius $R = 2\text{G}E$. Analogous result holds for $\eta=+1$, where the shell
is expanding: the apparent horizon is then in-going. The horizon radius is
determined by the equation $R_H = 2\text{G}E$, and the horizon will cut
$\Sigma$ if and only if $R_H > r$. We can assign the value $+1$ ($-1$) to the
horizon that is out- (in-)going and denote the quantity by $c$ (colour: black
or white hole). To summarize:
\begin{enumerate}
\item The condition that an apparent horizon intersects $\Sigma$ is
  $r < 2\text{G}E$.
\item The position of the horizon at $\Sigma$ is $R_H = 2\text{G}E$.
\item The colour $c$ of this horizon is $c = \eta$.
\end{enumerate}
 
In this way, questions about the existence and colour of an
apparent horizon outside the shell are reduced to equations containing
dynamical variables of the shell. In particular, the result that
$c=\eta$ can be expressed by saying that the shell always creates a
horizon outside that cannot block its motion. All that matters is that
the shell can bounce at the singularity (which it cannot within the
classical theory).

These results can be carried over to quantum mechanics after quantities such
as $2\mbox{G}E - r$ and $\eta$ are expressed in terms of the operators
describing the shell. Then we obtain a ``quantum horizon'' with the ``expected
radius'' $2\mbox{G}\langle E\rangle$ and with the ``expected colour''
$\langle\eta\rangle$ to be mostly black at the time when the expected radius
of the shell crosses the horizons inwards, neutrally grey at the time of the
bounce and mostly white when the shell crosses it outwards.

This proof has, however, two weak points. First, the spacetime metric on the
background manifold is not a gauge invariant quantity; although all gauge
invariant geometrical properties can be extracted from it within the classical
version of the theory, this does not seem to be possible in the quantum theory
\cite{paris}. Second, calculating the quantum spacetime geometry along
hyper-surfaces of a foliation on a given background manifold is foliation
dependent. For example, one can easily imagine two hyper-surfaces $\Sigma$ and
$\Sigma'$ belonging to different foliations, that intersect each other at a
sphere outside the shell and such that $\Sigma$ intersects the shell in its
in-going and $\Sigma'$ in its out-going state. Observe that the need for a
foliation is only due to our insistence on calculating the quantum metric on
the background manifold.

The essence of these problems is the gauge dependence of the results
of the calculation. However, it seems that this dependence concerns only
details such as the distribution of different hues of grey along the horizon,
not the qualitative fact that the horizon exists and changes colour from
almost black to almost white. Still, a more reliable method to establish the
existence and properties of grey horizons might require another material
system to be coupled to our model; this could probe the spacetime geometry
around the shell in a gauge-invariant way.

\subsection{Concluding remarks}
Comparison of the motion of wave packets of Sec.~\ref{sec:packets} with the
classical dynamics of the shell as described in Sec.~\ref{sec:mspace} shows a
notable difference. Whereas all classical shells cross their Schwarzschild
radius and reach the singularity in some stage of their evolution, the quantum
wave packets never reach the singularity, but always bounce and re-expand;
some of them even manage to cross their Schwarzschild radius during their
motion. This behaviour is far from being a small perturbation around a
classical solution if the classical spacetime is considered as a whole. Even
locally, the semi-classical approximation is not valid near the bouncing
point.  It is surely valid in the whole asymptotic region, where narrow wave
packets follow more or less the classical trajectories of the shell.

The most important question, however, concerns the validity of the
semi-classical approximation near the Schwarzschild radius. We have seen that
the geometry near the radius can resemble the classical black hole geometry in
the neighbourhood of the point where the shell is crossing the Schwarzschild
radius inwards. Then, the radius changes its colour gradually and the geometry
becomes very different from the classical one. Finally, near the point where
the shell crosses the Schwarzschild radius outwards, the radius is
predominantly white and the quantum geometry can be again similar to the
classical geometry, this time of a white hole horizon.

If the change of colour is very slow then the neighbourhood of the inward
crossing where the classical geometry is a good approximation can be large. It
seems that sufficiently large scattering time would allow for arbitrarily slow
change of colour. We cannot exclude, therefore, that the quantum spacetime
contains an extended region with the geometry resembling its classical
counterpart near a black hole horizon. This can be true even if the quantum
spacetime as a whole differs strongly from any typical classical collapse
solution.

One can even imagine the following scenario (which needs a more realistic
model than a single thin shell). A quantum system with a large energy
collapses and re-expands after a huge scattering time. The black hole horizon
phase is so long, that Hawking evaporation becomes significant and must be
taken into account in the calculation. It does then influence the scattering
time and the period of validity of the black hole approximation. The black
hole becomes very small and only then the change of horizon colour becomes
significant. The white hole stage is quite short and it is only the small
remnant of the system that, finally, re-expands. The whole process can still
preserve unitarity. In fact, this is a scenario for the issue of Hawking
evaporation process. It is not excluded by the results of the present paper.

One can also consider the following conception. At and under the Schwarzschild
radius, the local spacetime geometry for a white Schwarzschild radius is very,
and measurably, different from that of a black one. However, outside the
Schwarzschild radius, the local geometries of both cases are isometric to each
other so that the isometry need not contain the time reversal. Hence, the
quantum geometry outside a very grey horizon need not actually differ from the
classical geometry around a black hole (locally) very much.

All these speculations must be investigated and made more precise. If an
observer staying at the radius $R_B$ measures the proper time $\Delta T(R_B)$
between the shell wave packet crossing his position in and out, then of course
$\lim_{R_B \rightarrow \infty}\Delta T(R_B) = \infty$. In the collision
theory, one considers, therefore, some standard collision with $\Delta_s
T(R_B)$ and takes the limit only after the subtraction,
\[
  \lim_{R_B \rightarrow \infty}\bigl(\Delta T(R_B) - \Delta_s T(R_B)\bigr)\ .
\]
This limit, if finite, is called {\em time delay} \cite{dollard}. A
particularly difficult example is Coulomb scattering, for which the divergence
in $\Delta T(R_B)$ involves not only linear, but also logarithmic terms. This
is due to the long range of Coulomb potential and is, therefore, called {\em
  infrared divergence}. One can still find a subtraction for the Coulomb case
because the logarithmic term depends only on charges and so has the same
leading part for a whole class of scattering processes.

In our case, there are also logarithmic terms, but they depend on the energy of
the shell (energy is the charge for gravitation). No reasonable subtraction
seems, therefore, possible. One had better work at a finite radius $R_B$ all
time. To calculate our $\Delta T(R_B)$, one has to use the quantum geometry
outside the shell.

But what is the quantum geometry? In the classical version of general
relativity, a metric tensor field written in particular coordinates contains
all information about local geometric properties. A change in the
coordinates does not lead to any change of these properties because they can
be calculated from the transformed metric. In a quantum theory, however,
coordinate transformations is a delicate issue. Let us turn to our simple
system to see why. We have chosen the coordinates $U$ and $V$ that are
uniquelly determined for each solution by means of their geometric properties.
The components $A(U,V)$ and $R(U,V)$ of the four-metric with respect to these
coordinates are geometric quantities themselves. Nobody wants to deny that the
quantities $A(\eta,M,w;U,V)$ and $R(\eta,M,w;U,V)$ are Dirac observables for
each fixed value of $U$ and $V$: they are then just functions of $\eta$, $M$
and $w$, which {\em are} Dirac observables. In the quantum theory, one can try
to promote these quantities to operators by replacing $\eta$, $M$ and $w$ by
the corresponding quantum operators and by choosing a suitable factor
ordering. Let us suppose that the quantities $\hat{A}(U,V)$ and $\hat{R}(U,V)$
obtained in this way are well defined operators for each value of $U$ and $V$.
Cannot we calculate all geometric properties near each point $(U,V)$ from
these operators?

Problems emerge if we choose to work with a different set of geometric
coordinates. Let us, for example, pass to the Schwarzschild coordinates $T$
and $R$; near infinity, they are a very natural choice. Again, we can work
out the form of the metric components $g_{00}(\eta,M,w;T,R)$ and
$g_{11}(\eta,M,w;T,R)$ for each solution and try to construct the operators
$\hat{g}_{00}(T,R)$ and $\hat{g}_{11}(T,R)$ from them. Is there any
transformation within the quantum theory analogous to the classical coordinate
transformation and providing a basis for a proof that the local geometric
properties calculated from $\hat{A}(U,V)$ and $\hat{R}(U,V)$ are the same as
those calculated from $\hat{g}_{00}(T,R)$ and $\hat{g}_{11}(T,R)$? A
problem is that the transformation between $\{U,V\}$ and $\{T,R\}$ is field
dependent. Indeed, for $\eta = +1$, it is given by Eqs.~(\ref{Uinfty}) and
(\ref{Vinfty}). The right-hand sides depend not only on $T_\infty$ (which
coincides with our $T$ here) and $R$, but also on $M$ and $w$ (the full
transformation depends also on $\eta$). Hence, if we assume that the
quantities $T$ and $R$ are just parameters, then the quantities $U$ and $V$
are operators and vice versa. However, the quantities $T$ and $R$ are
parameters in the operators $\hat{g}_{00}(T,R)$ and $\hat{g}_{11}(T,R)$ and
the quantities $U$ and $V$ are parameters in $\hat{A}(U,V)$ and
$\hat{R}(U,V)$.

It seems, therefore, that there is a problem of principle in addition to the
factor-ordering problem to transform the operators $\hat{A}(U,V)$ and
$\hat{R}(U,V)$ to $\hat{g}_{00}(T,R)$ and $\hat{g}_{11}(T,R)$. The conclusion
is that quantum geometry cannot be described in a coherent manner analogous to
the differential geometry, see also \cite{paris}.

At the present stage, we try to calculate each geometric property for itself.
Thus, in his PhD thesis, M.~Ambrus, to whom I owe all my knowledge about
collision theory, tries to define and calculate the scattering times. Another
attempt is to define quantum geometry in a similar way as the classical
geometry is defined: by properties of test particles. This might work even
near the Schwarzschild and zero radius. Thus, we are trying with I.~Kouletsis,
to quantize an analogous system containing {\em two} shells and to use the
second shell as a spy probing the quantum geometry created by the first one.
For some preliminary results see \cite{H-Kou1}, \cite{H-Kou} and
\cite{H-Kou3}.

The calculations of this paper are valid only for null shells. Similar
calculations for massive shells have been performed in \cite{H-K-K}. There has
been re-expansion and unitarity for massive shells if the rest mass has been
smaller than the Planck mass ($10^-5$ g). It is very plausible that the
interpretation of these results is similar to that given in the present paper.
Thus, we can expect the results valid at least for all ``light'' shells. There
is, in any case, a long way to any astrophysically significant system and a
lot of work is to be done before we can claim some understanding of the
collapse problem.

\subsection*{Acknowledgments}
Discussions with M.~Ambrus, A.~Ashtekar, A.~Barvinski, R.~Beig, H.~Friedrich,
T.~Jacobson, J.~Jezierski, C.~Kiefer, J.~Kijowski, I.~Kouletsis,
K.~Kucha\v{r}, J.~Louko and C.~Vaz have strongly helped me to form the ideas
of these lectures.

\end{document}